# Neural Networks and Value at Risk

Alexander Arimond

Sociovestix Labs, Scotland, UK

Damian S. Borth

University of St. Gallen, Switzerland

Andreas G. F. Hoepner*

University College Dublin, Ireland, EU and

Technical Expert Group for Sustainable Finance, DG FISMA European Commission

Michael Klawunn

Warburg Invest AG, Germany, EU

Stefan Weisheit

Germany, EU

## ABSTRACT

Inspired by Gu, Kelly & Xiu's (GKX, 2020) advancement of the measurement of asset risk premia via the introduction of feed forward neural networks, we investigate, if machine learning can advance the process of 'estimating Value at Risk (VaR) thresholds'. For this purpose, we compare simple (GKX's feed forward) and advanced (convolutional, recurrent) neural networks with established approaches (Hidden Markov Model, Mean/Variance). Utilizing a generative regime switching framework, we perform Monte-Carlo simulations of asset returns for Value at Risk threshold estimation. Using equity markets and long term bonds as test assets in the global, US, Euro area and UK setting over an up to 1,250 weeks sample horizon ending in August 2018, we investigate neural networks along three design steps relating (i) to the initialization of the neural network, (ii) its incentive function according to which it has been trained and (iii) the amount of data we feed. First, we compare neural networks with random seeding with networks that are initialized via estimations from the best-established model (i.e. the Hidden Markov). We find latter to outperform in terms of the frequency of VaR breaches (i.e. the realized return falling short of the estimated VaR threshold). Second, we balance the incentive structure of the loss function of our networks by adding a second objective to the training instructions so that the neural networks optimize for accuracy while also aiming to stay in empirically realistic regime distributions (i.e. bull vs. bear market frequencies). In particular this design feature enables the balanced incentive recurrent neural network (RNN) to outperform the single incentive RNN as well as any other neural network or established approach by statistically and economically significant levels. Third, we half our training data set of 2,000 days. We find our networks when fed with substantially less data (i.e. 1,000 days) to perform significantly worse which highlights a crucial weakness of neural networks in their dependence on very large data sets. Hence, we conclude that well designed neural networks, i.e. a recurrent neural network initialized with best current evidence and balanced incentives – can potentially advance the protection offered to institutional investors by VaR thresholds through a reduction in threshold breaches. However, such advancements rely on the availability of a long data history, which may not always be available in practice when estimating asset management VaR thresholds.

Acknowledgments: We are grateful for comments from Theodor Cojoianu, James Hodson, Juho Kanniainen, Qian Li, Yanan, Andrew Vivian, Xiaojun Zeng and participants at 2019 Financial Data Science Association conference in San Francisco the International Conference on Fintech and Financial Data Science at University College Dublin (UCD). The views expressed in this manuscript are not necessarily shared by Sociovestix Labs, the Technical Expert Group of DG FISMA or Warburg Invest AG. Authors are listed in alphabetical order, whereby Hoepner serves as the contact author (andreas.hoepner@ucd.ie). Any remaining errors are our own.

# 1    Introduction

While leading papers on machine learning in asset pricing focus on predominantly returns and stochastic discount factors (Chen, Pelger & Zhu 2020; Gu, Kelly & Xiu 2020), we are motivated by the global Coid-19 virus crisis and the subsequent stock market crash to investigate if and how machine learning methods can enhance Value at Risk (VaR) threshold estimates. In line with Gu, Kelly & Xiu's (2020: 7), we like to open by disclaiming our awareness that "[m]achine learning methods on their own do not identify deep fundamental associations" .without human scientists designing hypothesized mechanisms into an estimation problem.[1] Nevertheless, measurement errors can be reduced based on machine learning methods. Hence, machine learning methods employed as means to an end instead of as end in themselves can significantly support researchers in challenging estimation tasks.[2]

In their already legendary paper, Gu, Kelly & Xiu (GKX in the following, 2020) apply Machine Learning to a key problem in academic finance literature: 'measuring asset risk premia'. They observe that machine learning improves the description of expected returns relative to traditional econometric forecasting methods based on (i) better out-of-sample R-squared and (ii) forecasts earning larger Sharpe ratios. More specifically, they compare four 'traditional' methods (OLS, GLM, PCR/PCA, PLS) with regression trees (e.g. random forests) and a simple 'feed forward neural network' based on 30k stocks over 720 months (1957-2016), using 94 firm characteristics, 74 sectors and 900+ baseline signals. Crediting inter alia (i) flexibility of functional form and (ii) enhanced ability to prioritize vast sets of baseline signals, they find the feed forward neural networks (FFNN) to perform best.

Contrary to results reported from computer vision, GKX further observe that "'shallow' learning outperforms 'deep' learning" (p.47), as their neural network with 3 hidden layers excels beyond neural networks with more hidden layers. They interpret this result as a consequence of a relatively much lower signal to noise ratio and much smaller data sets in finance. Interestingly, the outperformance of NNs over the other 5 methods widens at portfolio compared to stock level, another indication that an understanding of the signal to noise ratio in financial markets is crucial when training neural networks. That said, while classic OLS is statistically significantly weaker than all other models, NN3 beats all others but not always at statistically significant levels. GKX finally confirm their results via Monte Carlo simulations. They show that if one generated two hypothetical security price datasets, one linear and un-interacted and one nonlinear and interactive, OLS and GLM would dominate in former, while NNs dominate in the latter. They conclude by attributing the "predictive advantage [of neural networks] to accommodation of nonlinear interactions that are missed by other methods." (p.47)

Following GKX, an extensive literature on machine learning in finance is rapidly emerging. Chen, Pelger and Zhu (CPZ in the following, 2020) introduce more advanced (i.e. recurrent) neural networks and estimate a (i) non-linear asset pricing model (ii) regularized under no-arbitrage conditions operationalized via a stochastic discount factor (iii) while considering economic conditions. In particular they attribute the time varying dependency of the stochastic discount factor of about ten thousand US stocks to macroeconomic state processes via a recurrent Long Short Term Memory (LSTM) network. In CPZ's (2020: 5) view "it is essential to identify the dynamic pattern in macroeconomic time series before feeding them into a machine learning model".

---

[1] Furthermore, our research in the following does not aim to solve the significant challenges current machine learning methods face in terms of (i) replicability, (ii) technical interpretability and especially (iii) scientific explainability.

[2] In other words, 'Machine Learning' has seemingly infinite technical applications in empirical finance and may leapfrog or even retire some classic econometric methods. Hence, not exploring the potential benefits and risks of machine learning is not necessarily a progressive approach to academic research.

Avramov et al. (2020) replicate the approaches of GKX's (2020), CPZ (2020), and two conditional factor pricing models: Kelly, Pruitt, and Su's (2019) linear instrumented principal component analysis (IPCA) and Gu, Kelly, and Xiu's (2019) nonlinear conditional autoencoder in the context of real-world economic restrictions. While they find strong Fama French six factor (FF6) adjusted returns in the original setting without real world economic constraints, these returns reduce by more than half if microcaps or firms without credit ratings are excluded. In fact, when Avramov et al. (2020: 3) are "[e]xcluding distressed firms, all deep learning methods no longer generate significant (value-weighted) FF6-adjusted return at the 5% level." They confirm this finding by showing that the GKX (2020) and CPZ (2020) machine learning signals perform substantially weaker in economic conditions that limit arbitrage (i.e. low market liquidity, high market volatility, high investor sentiment). Curiously though, Avramov et al. (2020: 5) find that the only linear model they analyse - Kelly et al.'s (2019) IPCA – "stands out … as it is less sensitive to market episodes of high limits to arbitrage." Their finding as well as the results of CPZ (2020) imply that economic conditions have to be explicitly accounted for when analysing the abilities and performance of neural networks. Furthermore, Avramov et al. (2020) as well as GKX (2020) and CPZ (2020) make anecdotal observations that machine learning methods appear to reduce drawdowns.[1]

While their manuscripts focused on return predictability, we devote our work to risk predictability in the context of market wide economic conditions. The Covid-19 crisis as well as the density of economic crisis in the previous three decades imply that catastrophic 'black swan' type risks occur more frequent than predicted by symmetric economic distributions. Consequently, underestimating tail risks can have catastrophic consequences for investors. Hence, the analysis of risks with the ambition to avoid underestimations deserves, in our view, equivalent attention to the analysis of returns with its ambition to identify investment opportunities resulting from mispricing. More specifically, since a symmetric approach such as the "mean-variance framework implicitly assumes normality of asset returns, it is likely to underestimate the tail risk for assets with negatively skewed payoffs" (Agarwal & Naik, 2004:85). Empirically, equity market indices usually exhibit, not only since Covid-19, negative skewness in its return payoffs (Albuquerque, 2012, Kozhan et al. 2013). Consequently, it is crucial for a post Covid-19 world with its substantial tail risk exposures (e.g. second pandemic wave, climate change, cyber security) that investors provided with tools which avoid the underestimation of risks best possible. Naturally, neural networks with their near unlimited flexibility in modelling non-linearities appear suitable candidates for such conservative tail risk modelling that focuses on avoiding underestimations.

---

[1] Further recent applications of machine learning in finance include the work of Aminia et al. (2020) on capital structure, Aubry et al. (2019) on real assets, Bianchi et al. (2020) on bond return predictability, Easley et al. (2019) on microstructure, Götze et al. (2020) on catastrophe bonds, Hunt et al. (2019) on earnings forecasts, Sirignano et al. (2018) on mortgages, and Verstyuk (2019) on macroeconomic forecasts. Switching from application focused to more methodological work, dimension reduction techniques such as De Nard et al. (2020), Giglio & Xiu (2019), and Kozak, Nagel & Santosh (2020) as also noteworthy, as are efforts by Fallahgouly and Franstiantoz (2020) and Horel and Giesecke (2019) to develop significant tests for neural networks.

Our paper investigates is basic and/or more advanced neural networks have the capability of underestimating tail risk less often at common statistical significance levels. We operationalize tail risk as Value at Risk which is the most used tail risk measure in both commercial practice as well as academic literature (Billio et al. 2012, Billio and Pellizon, 2000, Jorion, 2005, Nieto & Ruiz, 2015). Specifically, we estimate VaR thresholds using classic methods (i.e. Mean/Variance, Hidden Markov Model)[1] as well as machine learning methods (i.e. feed forward, convolutional, recurrent), which we advance via initialization of input parameter and regularization of incentive function. Recognizing the importance of economic conditions (Avramov et al. 2020, Chen et al. 2020), we embed our analysis in a regime-based asset allocation setting.

Specifically, we perform Monte-Carlo simulations of asset returns for Value at Risk threshold estimation in a generative regime switching framework. Using equity markets and long term bonds as test assets in the global, US, Euro area and UK setting over an up to 1,250 weeks sample horizon ending in August 2018, we investigate neural networks along three design steps relating (i) to the initialization of the neural network's input parameter, (ii) its incentive function according to which it has been trained and which can lead to extreme outputs if it is not regularized as well as (iii) the amount of data we feed. First, we compare neural networks with random seeding with networks that are initialized via estimations from the best-established model (i.e. the Hidden Markov). We find latter to outperform in terms of the frequency of VaR breaches (i.e. the realized return falling short of the estimated VaR threshold). Second, we balance the incentive structure of the loss function of our networks by adding a second objective to the training instructions so that the neural networks optimize for accuracy while also aiming to stay in empirically realistic regime distributions (i.e. bull vs. bear market frequencies). This design features leads to better regularization of the neural network, as it substantially reduces extreme outcomes than can result from a single incentive function. In particular this design feature enables the balanced incentive recurrent neural network (RNN) to outperform the single incentive RNN as well as any other neural network or established approach by statistically and economically significant levels. Third, we half our training data set of 2,000 days. We find our networks when fed with substantially less data (i.e. 1,000 days) to perform significantly worse which highlights a crucial weakness of neural networks in their dependence on very large data sets.

Our contributions are fivefold. First, we extend the currently return focused literature of machine learning in finance (Avramov et al. 2020, Chen et al. 2020; Gu et al. 2020) to also focus on the estimation of risk thresholds. Assessing the advancements that machine learning can bring to risk estimation potentially offers valuable innovation to asset owners such as pension funds and can better protect the retirement savings of their members.[2] Second, we advance the design of our three types of neural networks by initializing their input parameter with the best established model. While initializations are a common research topic in core machine learnings fields such as image classification or machine translation (Glorot & Bengio, 2010, Zhang et al., 2019), we are not aware of any systematic application of initialized neural networks in the field of finance. Hence, demonstrating the statistical superiority of an initialized neural network over itself non-initialized appears a relevant contribution to the community. Third, while CPZ (2020) regularize their neural networks via no arbitrage conditions, we regularize via balancing the incentive function of our neural networks on multiple objectives (i.e. estimation accuracy and empirically realistic regime distributions). This prevents any single objective from leading to extreme outputs and hence balances the computational power of the trained neural network in desirable directions. In fact, our results show that amendments to the incentive function maybe the strongest tool available to us in engineering neural networks. Fourth, we also hope to make a marginal contribution to the literature on value at risk estimation. Whereas our paper is focused on advancing machine learning techniques and is therefore following Billio and Pellizon (2000) anchored in a regime based asset allocation setting[1] to account for time varying economic states (CPZ, 2020), we still believe that the nonlinearity and flexible form especially of recurrent neural networks maybe of interesting to the VaR (forecasting) literature (Billio et al. 2012, Nieto & Ruiz, 2015, Patton et al. 2019). Fifth, our final contribution lies in the documentation of weaknesses of neural networks as applied to finance. While Avramov et al. (2020) subjects neural networks to real world economic constraints and finds these to substantially reduce their performance, we expose our neural networks to data scarcity and document just how much data these new approaches need to advance the estimation of risk thresholds. Naturally, such long data history may not always be available in practice when estimating asset management VaR thresholds and therefore established methods and neural networks are likely to be used in parallel for the foreseeable future.

---

[1] While we acknowledge that hidden markov models and more generally (logistic) regressions can be seen as parametric machine learning methods themselves, if one uses a broader machine learning definition. We follow the GKX (2020) approach to machine learning here though and only refer to our neural networks as machine learning approaches.

[2] Lopez de Prado (2018) cites Cam Harvey's view that "[t]he first wave of quantitative innovation in finance was led by Markowitz optimization. Machine learning is the second wave and it will touch every aspect of finance." Following this view in the context of the Covid-19 crisis, it would be suitable if risk estimations were touched by machine learning sooner rather than later.

In section two, we will describe our testing methodology including all five competing models (i.e. Mean/Variance, Hidden Markov Model, Feed Forward Neural Network, Convolutional Neural Network, Recurrent Neural Network). Section three describes data, model training, Monte Carlo simulations and baseline results. Section four then advances our neural networks via initialization and balancing the incentive functions and discusses the results of both features. Section five conducts robustness tests and sensitivity analyses before section six concludes.

---

[1] We acknowledge that most recent statistical advances in Value at Risk estimation have concentrated on jointly modelling Value at Risk and Expected Shortfall and were therefore naturally less focused on time varying economic states (Patton et al. 2019, Taylor 2019, 2020).

## 2    Methodology

### 2.1    Value at Risk estimation with Mean/Variance approach

When modelling financial time series related to investment decisions the asset return $R_{tp}$ of portfolio (p) at time (t) as defined in equation (1) below is the focal point of interest instead of asset price $P_{tp}$, since investors earn on the difference between the price at which they sold.

$$R_{tp} = \frac{(P_{tp} - P_{t-1p})}{P_{t-1p}} \tag{1}$$

Value-at-Risk (VaR) metrics are an important tool in many areas of risk management. Our particular focus on VaR measures as a means to perform risk budgeting in asset allocation. Asset owners such as pension funds or insurances as well as asset managers often incorporate VaR measures into their investment processes (Jorion, 2005). Value at Risk is defined as in equation (2) as the lower bound of a portfolio's return, which the portfolio or asset is not expected to fall short off with a certain probability (a) within the next period of allocation (n).

$$\Pr\left(R_{t+np} < -VaR_{tp}(n)\right) = a \tag{2}$$

For example, an investment fund indicates that, based on the composition of its portfolio and on current market conditions, there is a 95% or 99% probability it will not lose more than a specified amount of assets over the next 5 trading days The VaR measurement can be interpreted as a threshold (Billio and Pellizon 2000). If the actual portfolio or asset return falls below this threshold, we refer to this a *VaR breach*.

The classic mean variance approach of measuring VaR values is based on the assumption that asset returns follow a (multivariate) normal distribution. VaR thresholds can then be measured by estimating the mean and covariance $(\mu, \Sigma)$ of the asset returns by calculating sample mean and sample covariance of the respective historical window. The 1% or 5% percentile of the resulting normal distribution will be an appropriate estimator of the 95% or 99% VaR threshold. We refer to this way of estimating VaR thresholds as being the "classical" approach and use it as baseline of our evaluation.

This classic approach, however, does not sufficiently reflect the skewness of real world equity markets and the divergences of return distributions across different economics regimes. In other words, the classic approach does not take into account longer term market dynamics, which express themselves as phases of growth or of downside, also commonly known as bull market and bear markets. For this purpose, regime switching models have grown in popularity well before machine learning entered finance (Billio and Pellizon 2000). In this study, we model financial markets inter alia using neural networks while accounting for shifts in economics regimes (Avramov et al. 2020, Chen et al., 2020). Due to the generative nature of these networks, they are able to perform Monte-Carlo simulation of future returns, which could be beneficial for VaR estimation.

## 2.2    Regime Switching with Hidden Markov Models

In asset manager's risk budgeting it is advantageous to know about the current market phase (regime) and estimate the probability that the regime changes (Schmeding et al., 2019). The most common way of modelling market regimes is by distinguishing between bull markets and bear markets. Unfortunately, market regimes are not directly observable, but are rather to be derived indirectly from market data. Regime Switching Models based on Hidden Markov Models are an established tool for regime based modelling. Hidden Markov Models (HMM) – which are based on Markov chains - are models that allow for analysing and representing characteristics of time series such as negative skewness (Ang and Bekaert, 2002; Timmerman, 2000). We employ the HMM for the special case of two economic states called 'regimes' in the HMM context.

Specifically, we model asset returns $y_t \in R^n$ (we are looking at $n \geq 1$ assets) at time $t$ to follow an $n$-dimensional Gaussian process with hidden states $S \in \{1, 2\}$ as shown in equation (3):

$$y_t \sim \mathcal{N}\big(\mu_{S_t}, \Sigma_{S_t}\big) \tag{3}$$

The returns are modelled to have state dependent expected returns $\mu_{S_t} \in R^n$ as well as covariance $\Sigma_{S_t} \in R^{nxn}$. The dynamic of $S_t$ is following a homogenous Markov chain with transition probability matrix

$$\begin{pmatrix} p & 1-p \\ 1-q & q \end{pmatrix}, 0 \leq p, q \leq 1 \tag{4}$$

with $p = P(S_t = 1 \mid S_{t-1} = 1)$ and $q = P(S_t = 2 \mid S_{t-1} = 2)$. This definition describes if and how states are changing over time. It is also important to note the 'Markov Property' that the probability of being in any state at the next point in time only depends on the present state, not the sequence of states that preceded it. Furthermore, the probability of being in a state at a certain point in time is given as $\pi_t = P(S_t = 1)$ and $(1 - \pi_t) = P(S_t = 2)$. This is also called smoothed state probability. By estimating the smoothed probability $\pi_T$ of the last element of the historical window as the present regime probability, we can use the model to start from there and perform Monte-Carlo simulations of future asset returns for the next $l$ days.[1] This is outlined for the two-regimes case in Figure 1 below.[2]

---

Figure 1: Algorithm for the Hidden Markov Monte-Carlo simulation (for two regimes)

| | |
|---|---|
| 1: Estimate $\phi = (\pi_0, A, \mu, \Sigma)$ from history $X_T$ | |
| 2: $p_0 \leftarrow \pi_T$ | ➤ Compute smoothed (regime) probability |
| 3: $s_0 \leftarrow S_t' \sim Bernoulli(p_0) + 1 \in \{1,2\}$ | ➤ draw first regime from Bernoulli distribution conditioned by $p_0$ |
| 4: for $i \in \{1, \dots, l\}$ do | |
| 5: $\quad p_i \leftarrow A_{s_{i-1}0}$ | ➤ Determine transition probability from previous regime |
| 6: $\quad s_i \leftarrow S_{t+i}' \sim Bernoulli(p_i) + 1$ | ➤ draw next regime |
| 7: $\quad X_{t+i}' \sim \mathcal{N}(\mu_{s_i}, \Sigma_{s_i})$ | ➤ draw return sample from regime's Gaussian |
| 8: end for | |

---

## 2.3 Neural Networks and Regime Switching Models

When Graves [13] successfully made use of a Long Short-Term Memory (LSTM) based recurrent neural network to generate realistic sequences of handwriting, he followed the idea of using a Mixture Density Network (MDN) to parametrize a Gaussian Mixture predictive distribution (Bishop, 1995). Compared to standard neural networks (Multi-Layer Perceptron) as used by GKX (2020), this network does not only predict the conditional average of the target variable as point estimate (in GKX' case expected risk premia), but rather estimates the conditional distribution of the target variable. Given the autoregressive nature of Graves' approach, the output distributions are not assumed to be static over time, but dynamically conditioned on previous outputs, thus capturing the temporal context of the data. We consider both characteristics as being beneficial for modelling financial market returns, which experience a low signal to noise ratio as highlighted by GKX' results due to inherently high levels of intertemporal uncertainty.

---

[1] The parameter of the model $\varphi = (\pi_0, A, \mu, \Sigma)$ can be estimated based on historical return data of some window size. This is done by using the Baum-Welch algorithm, which is an expectation-maximization (EM) algorithm.
[2] It is worth noting that for the one regime, the algorithm estimates a standard multivariate Gaussian distribution and therefore mimics the classic Mean/Variance method.

The core of the proposed neural network regime switching framework is a (swappable) neural network architecture, which takes as input the historical sequence of daily asset returns. At the output level, the framework computes regime probabilities and provides learnable gaussian mixture distribution parameters, which can be used to sample new asset returns for Monte-Carlo simulation. A multivariate gaussian mixture model (GMM) is a weighted sum of $k$ different components, each following a distinct multivariate normal distribution as shown in equation (5):

$$P(X) = \sum_{i=0}^{k} \phi_i \mathcal{N}(X \mid \mu_i, \Sigma_i) \text{ with } \sum_{i=0}^{k} \phi_i = 1 \tag{5}$$

A GMM by its nature does not assume a single normal distribution, but naturally models a random variable as being the interleave of different (multivariate) normal distributions. In our model, we interpret $k$ as the number of regimes and $\varphi_i$ explains how much each regime contributes to the (current output). In other words, $\varphi_i$ can be seen as the probability that we are in regime $i$. In this sense the GMM output provides a suitable level of interpretability for the use case of regime based modelling. With regard to the neural network regime switching model, we extend the notion of a gaussian mixture by conditioning $\varphi_i$ via a yet undefined neural network f on the historic asset returns within a certain window of a certain size. We call this window *receptive field* and denote its size by $r$:

$$\phi_i(t) = f(X^{t-r,t}) = P(\phi_i \mid X_{t-r}, \ldots, X_t) \tag{6}$$

This extension makes the gaussian mixture weights dependent on the (recent) history of the time varying asset returns. Note that we only condition $\varphi$ on the historical returns. The other parameters of the gaussian mixture $(\mu_i, \Sigma_i)$, are modelled as unconditioned, yet optimizable parameters of the model. This basically means we assume the parameters of the gaussians to be constant over time (per regime). This is in contrast to the standard MDN, where $(\mu_i, \Sigma_i)$ are also conditioned on $X$ and therefore can change over time.[1] Keeping these remaining parameters unconditional is crucial to allow for a fair comparison between the neural networks and the HMM, which also exhibits time invariant parameters $(\mu_i, \Sigma_i)$ in its regime shift probabilities. Following Graves (2013), we define the probability given by the network and the corresponding sequence as shown in equation (7) and (8), respectively:

$$P(X) = \prod_{t=1}^{T} P(X_{t+1} \mid \phi(t), \mu, \Sigma) \tag{7}$$

$$\mathcal{L}(X) = -\sum_{t=1}^{T} \log P(X_{t+1} \mid \phi(t), \mu, \Sigma) \tag{8}$$

---

[1] We found this structure to be hard to interpret from a regime modelling point of view: If the regime distributions change at each point of time, it is hard to infer whether a regime describes a bullish or bearish market in general. With fixed optimizable distribution parameters we can interpret the distribution as belonging to either a bull market regime (positive expected return, low volatility) or bear market regime (negative expected return, high volatility).

Since financial markets operate in weekly cycles with many investors shying away from exposure to substantial leverage during the illiquid weekend period, we are not surprised to observe that model training is more stable when choosing the predictive distribution to not only be responsible for the next day, but for the next 5 days (Hann and Steuer, 1995). We call this forward looking window the *lookahead*. This is also practically aligned with the overall investment process, in which we want to appropriately model the upcoming allocation period, which usually spans multiple days. It also fits with the intuition that regimes do not switch daily but have stability at least for a week. The extended sequence probability and sequence loss are denoted accordingly in equations (9) and (10):

$$P(X) = \prod_{t=1}^{T} \prod_{j=0}^{5} P\big(X_{t+j}\big|\phi(t), \mu, \Sigma\big) \tag{9}$$

$$\mathcal{L}(X) = -\sum_{t=1}^{T} \sum_{j=0}^{5} \log P\big(X_{t+j}\big|\phi(t), \mu, \Sigma\big) \tag{10}$$

An important feature of the neural network regime model is how it simulates future returns. We follow Graves (2013) approach and conduct sequential sampling from the network. When we want to simulate a path of returns for the next $N$ business days, we do this according to the algorithm displayed in Figure 2.

---

Figure 2: : Algorithm for Neural Network Regime Switching Model - Monte-Carlo simulation (2-regime case)

1:    Train model on history $X_T$

2:    for $i \in \{0, \dots, l - 1\}$ do

3:        $\phi(t + i + 1) \leftarrow \phi(X^{t-r+i, t+i})$       ➢  Apply model on receptive field

4:        $r_{i+1} \leftarrow R'_{t+i+1} \sim B\big(\phi_0(t + i + 1)\big) + 1 \in \{1,2\}$   ➢  draw sample regime as conditioned by a Bernoulli distribution

5:        $X'_{t+i+1} \sim \mathcal{N}\big(\mu_{r_{i+1}}, \Sigma_{r_{i+1}}\big)$       ➢  draw return sample from regime Gaussian

6:        Append $X'_{t+i+1}$ $to$ $X$

7:    end for

---

We display our research design in Figure 3 'Neural Network Regime Switching Model'. The network takes as input the most recent days (of receptive field size r). The Temporal Neural Network block is interchangeable (e.g. simple feed forward, CNN, LSTM). The network conditions are the regime probabilities of the Gaussian Mixture Model (GMM). The residual parameters are unconditioned, learnable parameters of the model. The network is trained by targeting the next 5 days. This is called lookahead. As building block for the temporal neural network part of the model we choose three different neural network architectures, which we introduce in the following sections.

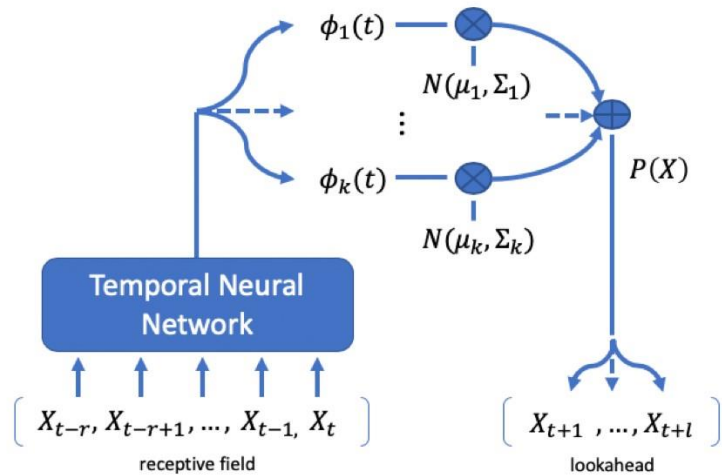

Figure 3: Neural Network Regime Switching Model

## 2.4 Feed Forward Neural Network

In accordance with GKX (2020) we first focus our analysis on traditional "feed-forward" neural networks before engaging in more sophisticated neural network architectures for time series analysis within the neural network regime model. The traditional model of neural networks, also called Multi-Layer Perceptron, consists of an "input layer" which contains the raw input predictors and one or more "hidden layers" that combine input signals in a nonlinear way and an "output layer", which aggregates the output of the hidden layers into a final predictive signal. The nonlinearity of the hidden layers arises from the application of nonlinear "activation functions" on the combined signals. We visualise the traditional feed forward neural network and its input layers in Figure 4.

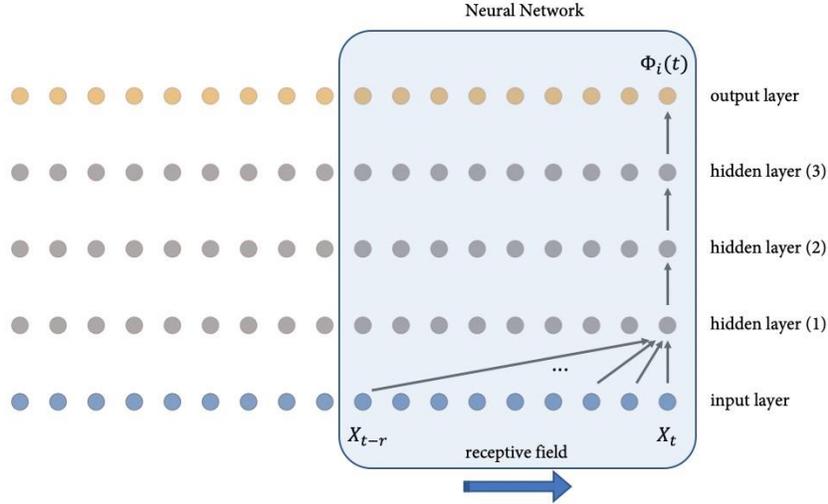

Figure 4: Feed Forward Neural Network

The figure depicts how the networks reads the data as a sequence from left to right. At a single point in time, the network takes as input the asset returns of the last $N = 10$ days. In each layer, there is fixed number of hidden units (32, 16, 8) which are not visualized here. In between layers it uses tanh as activation function. The output layer aggregates the hidden layer output via SoftMax into regime probabilities.

we setup our network structure in alignment with GKX's (2020) best performance neural network 'NN3'. The setup of our network is thus given with 3 hidden layers with decreasing number of hidden units (32, 16, 8). Since we want to capture the temporal aspect of our time series data, we condition the network output on at least a receptive field of 10 days. Even though the receptive field of the network is not very high in this case, the dense structure of the network results in a very high number of parameters (1698 in total, including the GMM parameters). In between layers, we make use of the activation function *tanh*.

## 2.5   Temporal Convolutional Neural Networks

Convolutional Neural Networks (CNNs) can also be applied within the proposed neural network regime switching model. Recently, CNNs gained popularity for time series analysis, as for example Van den Oord et al. (2015) successfully applied convolutional neural networks on time series data for generating audio waveforms, the state-of-the-art text-to-speech and music generation. Their adaption of Convolutional Neural Networks – called WaveNet – has shown to be able to capture long ranging dependencies on sequences very well. In its essence, a WaveNet consists of multiple layers of stacked convolutions along the time axis. Crucial features of these convolutions are that they have to be causal and dilated. Causal means that the output of a convolution only depends on past elements of the input sequence. Dilated convolutions are ones that exhibit "holes" in their respective kernel, which effectively means that its filter size increases while being dilated with zeros in between. WaveNet typically is constructed with increasing dilation factor (doubling in size) in each (hidden) layer. By doing so, the model is capable of capturing an exponentially growing number of elements from the input sequence depending on the number of hidden convolutional layers in the network. The number of captured sequence elements is called the receptive field of the network (and in this sense is equal to the receptive field defined for the neural network regime model).[1]

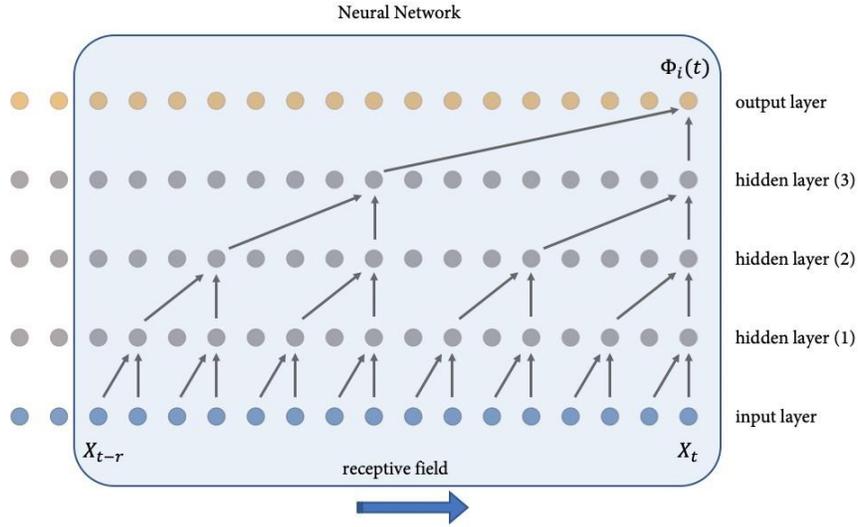

Figure 5: Visualization of Causal and Dilated Convolution as in Wavenet

The Convolutional Neural Network (CNN), due to its structure of stacked dilated convolutions, has a much greater receptive field than the simple feed forward network and needs much less weights to be trained. We restricted the number of hidden layers to 3 to illustrate the idea. Our network structure has 7 hidden layers. Each hidden layer furthermore exhibits a number of channels, which are not visualized here.

Figure 5 illustrates the networks basic structure as a combination of stacked causal convolutions with a dilation factor of $D = 2$. The backing model presented in this investigation is inspired by WaveNet, We restrict the model to the basic layout, using causal structure and increasing dilation between layers. The output layer comprises the regime predictive distributions by applying a SoftMax function to the hidden layers' outputs. Our network consists of 6 hidden layers, each layer having 3 channels. The convolutions each have a kernel size of 3. In total, the network exhibits 242 weights (including GMM parameters), the receptive field has a size of 255 days.

## 2.6    Long Short-Term Memory (LSTM) Recurrent Neural Network

As Graves (2013) was very successful in applying LSTM for generating sequences, we also adapt this approach for the neural network regime switching model. Originally introduced by Hochreiter and Schmidhuber (1997), a main characteristic of LSTMs – which are a sub class of recurrent neural networks - is its purpose-built memory cells, which allows it to capture long range dependencies in the data. From a model perspective,  LSTMs differ from other neural network architectures in that they are applied recurrently (see Figure 6).

---

[1] Van den Oord et al. further describe activation functions and skip connections for WaveNet. For the interested reader, we kindly refer to their paper for an in-depth description.

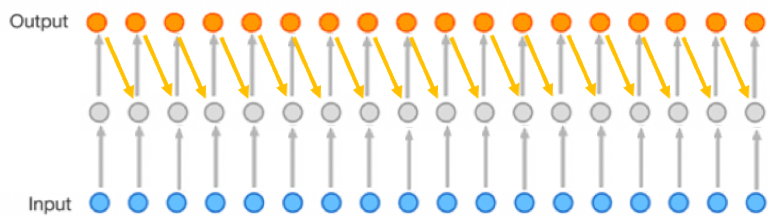

Figure 4: Temporal Dynamic of a Recurrent Neural Network

The Recurrent Neural Network (RNN) is characterized by its feedback loop: the output of a previous iteration of the function is used additionally as input when reading the next sequence element. Long Short-Term Memory (LSTM), which are a special variant of RNN, are able to capture long range sequence dependencies in this way.

The output from a previous sequence of the network function serves – in combination with the next sequence element - as input for the next application of the network function. In this sense, the LSTM can be interpreted as being similar to an HMM, in that there is a hidden state which conditions the output distribution. However, the LSTM hidden state not only depends on its previous states, but it also captures long term sequence dependencies through its recurrent nature. Maybe most notably, the receptive field size of an LSTM is not bound architecture wise as in case of simple feed forward network and CNN. Instead, the LSTM's receptive field depends solely on the LSTMs ability to memorize the past input. In our architecture we have one LSTM layer with a hidden state size of 5. In total, the model exhibits 236 parameters (including the GMM parameters). The potential of LSTMs was noted by CPZ (2020: 6) who note that "LSTMs are designed to find patterns in time series data and … are among the most successful commercial AIs".

## 3    Assessment Procedure

### 3.1    Data

We obtain daily price data for stock and bond indices globally for three major global markets (i.e. EU, UK, US) to study the presented regime based neural network approaches on a variety of stock markets and bond markets. For each stock market, we focus on one major stock index. For bond markets, we further distinguish between long term bond indices (7-10 years) and short term bond indices (1-3 years). The markets in scope are (1) USA, represented by S&P 500 and US Treasury Bonds, (2) Europe, represented by EURO STOXX 50 and German Bundesanleihen, as well as (3) United Kingdom with FTSE 100 and UK government bonds. Furthermore, we look at the (4) global market by looking at MSCI World Index in cross-section to US Treasuries as being the most globally important government bonds. Each model is trained includes equities, short bonds and long bonds (i.e. $X_t \in R^3$).

The data dates back to at least January 1990 and ends with August 2018, which means covering almost 30 years of market development. Hence, the data also accounts for crises like the dot-com bubble in the early 2000s as well as the financial crisis of 2008. This is especially important for testing the regime based approaches. The price indices are given as total return indices (i.e. dividends treated as being reinvested) to properly reflect market development. The data is taken from Refinitiv's DataStream.

Descriptive statistics are displayed in Table 1, whereby Panel A displays a daily frequency and Panel B a weekly frequency. Mean returns for equities exceed the returns for bond whereby the longer bond return more than the shorter one. Equities have naturally a much higher standard deviation and a far worse minimum return. In fact, equity returns in all four regions lose substantially more money than bond return even at the $25^{th}$ percentile, which highlights that the holy grail of asset allocation is the ability to predict equity market drawdowns. Furthermore, equity markets tend to bequite negatively skewed as expected while short bonds experience a positive skewness, which reflects previous findings (Albuquerque, 2012, Kozhan et al. 2013) and the inherent differential in the riskiness of both asset's payoffs.

[Insert Table 1 about here]

### 3.2    Model Training

The back testing is done on a weekly basis via a moving window approach. At each point in time, the respective model is fitted by providing the last $2,000$ days (which is roughly 8 years) as training data. We choose this long range window, because neural networks are known to need big datasets as inputs and it is reasonable to assume that over eight years include simultaneously times of (at least relative) crisis and times of market growth. Covering both bull and bear markets in the training sample is crucial to allow the model to "learn" these types of regimes.[1] For all our models we set the number of regimes to $k = 2$. As we back test an allocation strategy with a weekly re-allocation, we set the lookahead for the neural network regime models to 5 days. We further configured the back testing dates to always align with the end of a business week (i.e. Fridays).

The Classic approach does not need any configuration, model fitting is same as computing sample mean and sample covariance of the asset returns within the respective window. The HMM also does not need any more configuration, the Baum-Welch algorithm is guaranteed to converge the parameters into a local optimum with respect to the likelihood function (Baum, 1970).

For the neural network regime models, additional data processing is required to learn network weights that lead to meaningful regime probabilities and distribution parameters. An important pre-processing step is input normalization, as it is considered good practice for neural network training (Bishop, 1995). For this purpose, we normalize the input data by $X' = (X - mean(X)) / var(X)$. In other words, we demean the input data and scale them by their variance but without removing the interactions between the assets. We train the network by using the AdaMax optimizing algorithm (Kingma & Ba, 2014) and at the same time applying weight decay to reduce overfitting (Krogh & Hertz, 1992). Learning rate and number of epochs configured for training vary depending on the model.

In general, estimating parameters of a neural network model is a non-convex optimization problem. Thus, the optimization algorithm might become stuck in an infeasible local optimum. In order to mitigate this problem, it is common practice to repeat the training multiple times, starting off having different (usually randomly chosen) parameter initializations, and then averaging over the resulting models or picking the best in terms of loss. In this paper, we follow a best-out-of-5 approach, that means each training is done five times with varying initialization and the best one is selected for simulation. The initialization strategy, which we will show in chapter 4.1, further mitigates this problem by starting off from an economically reasonable parameter set.

We observe that the in-sample regime probabilities learned by the neural network regime switching models as compared to those estimated by the HMM based regime switching model generally show comparable results in terms of distribution and temporal dynamics. When we set $k = 2$ and the model fits two regimes with nearly invariably one having a positive corresponding equity means and low volatility, and the other experiencing a low or negative equity mean and high volatility. These regimes can be interpreted as bull and bear market, respectively. The respective in-sample regime probabilities over time also show strong alignment with growth and drawdown phases. This holds true for the vast majority of seeds and hence indicates that the neural network regime model is a valid practical alternative for regime modelling when compared to a Hidden Markov Model.

---

[1] We note that learning following Confucius includes imitation, experience and reflecting, whereas machine learning uses the term largely to refer to the former.

### 3.3    Simulation

After training the model for a specific point in time, we start a Monte Carlo simulation of asset returns for the next 5 days (one week - Monday to Friday). For the purpose of calculating statistically solid quantiles of the resulting distribution, we simulate 100,000 paths for each model. We do this for at least 1093 (EMU), and at most 1250 (globally) points in time within the back-test history window. As soon as we have simulated all return paths, we calculate a total (weekly) return for each path. The generated weekly returns follow a non-trivial distribution, which arises from the respective model and its underlying temporal dynamics. Based on the simulations we compute quantiles for Value at Risk estimations. For example, the 0.01 and 0.05 percentile of the resulting distribution represent the 99% and 95% - 5 day – VaR metric, respectively.

We evaluate the quality of our Value at Risk estimations by counting the number of breaches of the asset returns. In case, the actual return is below the estimated VaR threshold, we count this as a breach. Assuming an average performing model, it is e.g. reasonable to expect 5% breaches for a 95% VaR measurement.

We compared the breaches of all models with each other. We classify a model as being superior to another model, if the number of VaR breaches is less than those from the compared model. A value comparison *comp* = 1.0(= 0.0) indicates that the row model is superior (inferior) to the column model. We performed significance tests by applying paired t-tests.  We further evaluated a *dominance* value which is defined as shown in equation (11):

$$dom(model1|model2) = \frac{breaches(model1) \cap breaches(model2)}{breaches(model1)} - 1 \qquad (11)$$

A value *dom* = 0.0 means that any VaR breach that occurs with model 1 also occurs with model 2. A negative value means that model 1 exhibits VaR breaches that do not occur with model 2. In that sense, model 2 is dominating model 2 for this particular breach. The lower the dom value the more cases in which model 1 has a breach that model 2 does not share. If the dom value, however, is zero and model 1 wins the comparison against model 2 (i.e. comp = 1.0), then model 1 fully dominates model 2, as it has less breaches and every time it breaches, model 2 does too.

## 4        Discussion of Results by Design Feature

In our view the three most crucial design features of neural networks in finance, where the sheer number of hidden layers appears less helpful due to the low signal to noise ratio (GKX, 2020), are: amount of input data, initializing information and incentive function.

Big input data is important for neural networks, as they need to consume sufficient evidence also of rarer empirical features to ensure that their nonlinear abilities in fitting virtually any functional form are used in a relevant instead of an exotic manner. Similarly, the initialization of input parameters should be as much as possible based on empirically established estimates to ensure that the gradient descent inside the neural network takes off from a suitable point of departure, thereby substantially reducing the risks that a neural network confuses itself into irrelevant local minima.

On the output side, every neural network is trained according to an incentive (i.e. loss) function. It is this particular loss function which determines the direction of travel for the neural network, which has no other ambitions than to minimize its loss best possible. Hence, if the loss function only represents one of several practically relevant parameters, the neural network may come to results with bizarre outcomes for those parameters not included in its incentive function. In our case, for instance, the baseline incentive is just estimation accuracy which could lead to forecasts dominated much more by a single regime than ever observed in practice. In other words, after a long bull market, the neural network could "conclude" that bear markets do not exist. Metaphorically spoken, a unidimensional loss function in a neural network has little decency (Marcus, 2018).

Commencing with the initialization and the incentive functions, we will assess our three neural networks in the following vis a vis classic and HMM approach, where each of the three networks is once displayed with an advanced design feature and once with a naïve design feature.

### 4.1        Initialization

If no specific initialization strategy for neural networks is defined, it occurs entirely random, normal via a computer generated random number. Where established econometric approaches use naïve priors (i.e. mean), neural networks originally relied on brute force computing power and a bit of luck. Hence, it is unsurprising that initializations are a common research topic in core machine learnings fields such as image classification or machine translation (Glorot & Bengio, 2010, Zhang et al., 2019) nowadays. However, we are not aware of any systematic application of initialized neural networks in the field of finance. Hence, we compare naïve neural networks, which are not initialized with neural networks that have been initialized with the best available prior. In our case, the best available prior for $\mu_i, \Sigma_i$ of the model is the equivalent HMM estimation based on the same window.[1] Such initialization is feasible, since the structure of the Neural Network - due to its similarity with respect to $\mu_i, \Sigma_i$ – is broadly comparable with the HMM. In other words, we make use of already trained parameters from HMM training as starting parameters for the Neural Network training. In this sense, initialized neural networks are not only flexible in their functional form, they are also adaptable to "learn" from the best established model in the field if suitably supervised by the human data scientists. Metaphorically spoken, our neural networks can stand on the shoulders of the giant that HMM is for regime based estimations.

Table 2 presents the results by comparing breaches between the two classic approaches (Mean/Variance, HMM) and the non-initialized and HMM initialized neural networks across all four regions. Panel A and B display the 1% VaR threshold for equities and long bonds, respectively, while Panels C and D show the equivalent comparison for 5% VaR thresholds.[2] Note that for model training we apply a best-out-of-5 strategy as described in section 3.2. That means we repeat the training five times, starting off with random parameter initializations each time. In case of the presented HMM initialized model, we apply the same strategy, with the exception that $\mu_i, \Sigma_i$ of the model are initialized the same for each of the five iterations. All residual parameters are initialized randomly as fits best according to the neural network part of the model. XXX findings are observable:

First, not a single VaR threshold estimation process in a single region and in either of the two asset classes was able uphold its promise in that an estimated 1% VaR threshold should be breached no more than 1% of the time. This is very disappointing and quite alarming for institutional investors such as pension funds and insurance since it implies that all approaches – established and machine learning based – fail to sufficiently capture downside tail risks and hence underestimate 1% VaR thresholds. The vast majority of approaches estimate VaR thresholds that occur in more than 2% of the cases and the LSTM fails entirely if not initialised. In fact, even the best method, the HMM for US equities, estimates VaR thresholds which are breached in 1.34% of the cases.

Second, when inspecting the ability of our eight methods to estimate 5% VaR thresholds, the result remains bad but is less catastrophic. The Mean/Variance approach, the HMM and the initialised LSTM display cases where their VaR thresholds were breaches in less than the expected 5%. The Mean/Variance and HMM approach make their thresholds in 3 out of 8 cases and the initialised LSTM in 1 out of 8. Overall, this is still a disappointing performance, especially for the feed forward neural network and the CNN.

---

[1] Even though we initialize $\mu_i, \Sigma_i$ from HMM parameters, we still have weights to be initialized arising from the temporal Neural Network part of the model. We do this on a per layer level by sampling uniformly as $w \leftarrow \mathcal{U}(-\frac{1}{\sqrt{i}}, \frac{1}{\sqrt{i}})$ where $i$ is the number of input units for this layer.

[2] We focus our discussion of results on the equities and long bonds since these have more variation, lower skewness and hence risk. Results for the short bonds are available upon request from the contact author.

Third, when comparing the initialised with the non-initialised neural networks, the performance is like day vs. night. The non-initialised neural networks perform always worse and the LSTM performs entirely dismal without a suitable prior. When comparing across all eight approaches, the HMM appears most competitive which means that we either have to further advance the design of our neural networks or their marginal value add beyond classic econometric approaches appears inexistent. To advance the design of our neural networks further, we aim to balance its utility function to avoid extreme unrealistic results possible in the univariate case.

[Insert Table 2 about here]

## 4.2   Balancing incentive functions

Whereas CPZ (2020) regularize their neural networks via no arbitrage conditions, we regularize via balancing the incentive function of our neural networks on multiple objectives. Specifically, we extend the loss function to not only focus on accuracy of point estimates but also give some weight to eventually achieving empirically realistic regime distributions (i.e. in our data sample across all four regions no regimes display more than 60% frequency on a weekly basis). This balanced extension of the loss function prevents the neural networks from arriving at bizarre outcomes such as the conclusion that bear markets (or even bull markets) barely exist.

Technically, such bizarre outcomes result from cases where the regime probabilities $\varphi_i(t)$ tend to converge globally either into 0 or 1 for all $t$, which basically means the neural network only recognises one-regime. To balance the incentive function of the neural network and facilitate balancing between regime contributions, we introduced an additional regularization term $reg$ into the loss function which penalizes unbalanced regime probabilities. The regularization term is displayed in equation (13) below. If bear and bull market have equivalent regime probabilities the term converges to 0.5, while it converges towards 1 the larger the imbalance between the two regimes.

$$reg(x) = \sum_{i=1}^{k} \left( \frac{1}{T} \sum_{t=1}^{T} \phi_i(t) \right)^2 = \sum_{i=1}^{k} \overline{\phi_i(t)}^2 \tag{13}$$

Substituting equation (13) into our loss function of equation (10), leads to equation (14) below, which doubles the point estimation based standard loss function in case of total regime balance inaccuracy but adds only 50% of the original loss function in case of full balance. Conditioning the extension of the loss function on its origin is important to avoid biases due to diverging scales. Setting the additional incentive function to initially have half the marginal weight of the original function also seems appropriate for comparability.

$$\tilde{\mathcal{L}}(x) = \big(1 + reg(x)\big)\mathcal{L}(x) \tag{14}$$

The outcome of balancing the incentive functions of our neural networks are displayed in Table 3, where Panels A-D are distributed as previously in Table 2. The results are very encouraging, especially with regards to the LSTM. The regularized LSTM is in all 32 cases (i.e. 2 thresholds, 2 asset classes, 4 regions) better than the non-regularized LSTM. For the 5% VaR thresholds, it reaches realized occurrences of less than 4% in half the cases. This implies that the regularized LSTM can even be more cautious than required. The regularized LSTM also sets a new record for the 1% VaR thresholds with only 1.22% breaches for long UK bonds but all eight approaches remain to underestimate the downside tail risk with their VaR threshold estimations. The standard feed forward neural network also enhances its performance following the incentive balancing regularization in nearly all cases while the CNN regularization delivers a more mixed picture

[Insert Table 3 about here]

Table 4 displays the direction comparisons between the eight approaches for all four regions with the 1% (5%) VaR threshold results being displayed for equities and Long Bonds in Panels A and B (C and D), respectively. The regularized LSTM outperforms all other approaches at statistically significant levels across virtually all contexts. Only the HMM and to a lesser extend the regularized FNN can occasionally get into statistically indifferent territory and the HMM, in exceptions, manages to produce lesser breaches. In fact, the regularized LSTM substantially dominates especially the classic mean/variance approach and the non-regularized neural networks in many contexts. This result implies that HMM initialised, regularized LSTMs can add real value to the risk management process of institutional investors such as pension funds and insurance.

[Insert Table 4 about here]

To measure how much value the regularized LSTM can add compared to alternative approaches, we compute the annual accumulated costs of breaches as well as the average cost per breach. They are displayed in Table 5 for the 5% VaR threshold. The regularized LSTM is for both numbers in any case better than the classic approaches (Mean/Variance ad HMM) and the difference is economically meaningful. For equities the regularized LSTM results in annual accumulated costs of 97-130 basis points less than the classic Mean/Variance approach, which would be up to over one billion US$ avoid loss per annum for a > US$100 billion equity portfolios of pension fund such as CalPERS or PGGM. Compared to the HMM approach, the regularized LSTM avoids annual accumulated costs of 44-88 basis points, which is still a substantial amount of money for the vast majority of asset owners. With respect to long bonds, where total returns are naturally lower, the regularized LSTM's avoided annual costs against the mean/variance and the HMM approach range between 23-30 basis points, which is high for bond markets.

[Insert Table 5 about here]

## 4.3   Size of Input Data

These statistically and economically attractive results have been achieved, however, based on 2,000 days of training data. Such "big" amounts of data may not always be available for newer investment strategies. Hence, it is natural to ask if the performance of the regularized neural networks drop when fed with just half the data (i.e. 1,000 days). Apart from reducing statistical power, a period of over 4 years also may comprise less information on downside tail risks. Indeed, the results displayed in Table 6 show that in all context of VaR thresholds and asset classes, the regularized networks trained on 2,000 days substantially outperform and usually dominate their equivalently designed neural networks with half the training data. Hence, the attractive risk management features for HMM initialised, balanced incentive LSTMs are likely only available for established discretionary investment strategies where sufficient historical data is available or for entirely rules-based approaches whose history can be replicated ex-post with sufficient confidence.

[Insert Table 6 about here]

## 5       Robustness Tests and Sensitive Analysis[1]

We further conduct an array of robustness tests and sensitivity analysis to challenge our results and the applicability of neural network based regime switching models. As first robustness test, we extend the regularization in a manner that the balancing incentive function of equation (13) has the same marginal weight than the original loss function instead of just half the marginal weight. The performance of both types of regularized LSTMs is essentially equivalent Second, we study higher VaR thresholds such as 10% and find the results to be very comparable to the 5% VaR results. Third, we estimate monthly instead of weekly VaR. Accounting for the loss of statistical power in comparison tests due to the lower number of observations, the results are equivalent again.

We conduct two sensitivity analysis. First, we set up our neural networks to be generalized by two balancing incentive functions but without HMM initialisation. The results show the regularization enhances performance compared to the naïve non-regularized and non-initialized models but that both design features are needed to achieve the full performance. In other words, initialization and regularization seem additive design features in terms of neural network performance. Second, we run analytical approaches with K > 2 regimes. Adding a third or even fourth regime when asset prices only know two directions leads to substantial instability in the neural networks and tends to depreciate the quality of results.

---

[1] Results are available upon request from the contact author.

## 6 Conclusion

Inspired by GKX (2020)'s and CPZ (2020)'s recent research into machine learning in finance and the current renewed focus on tail risks during the Covid-19 crisis, we investigate in this manuscript if neural networks can be beneficial for Value at Risk threshold estimation. We introduced a framework architecture which allows for learning of regime switching models based on neural networks. By doing so, we were able to apply and evaluate state-of-the-art temporal neural network models (CNN and LSTM) in the domain of regime switching and VaR estimation. Utilizing our generative regime switching framework, we perform Monte-Carlo simulations of asset returns for Value at Risk threshold estimation. Employing equity markets and long term bonds as test assets in the global, US, Euro area and UK setting over an up to 1,250 weeks sample horizon ending in August 2018, we investigate neural networks along three design steps.

First, we compare neural networks with random seeding with networks that are initialized via estimations from the best established model (i.e. the Hidden Markov). We find latter to outperform in terms of the frequency of VaR breaches (i.e. the realized return falling short of the estimated VaR threshold). Second, we balance the incentive structure of the loss function of our networks by adding a second objective to the training instructions so that the neural networks optimize for accuracy while also aiming to stay in empirically realistic regime distributions (i.e. bull vs. bear market frequencies). In particular this design feature enables the balanced incentive recurrent neural network (RNN) to outperform the single incentive RNN as well as any other neural network or established approach by statistically and economically significant levels. Third, we half our training data set of 2,000 days. We find our networks when fed with substantially less data (i.e. 1,000 days) to perform significantly worse which highlights a crucial weakness of neural networks in their dependence on very large data sets.

Hence, we conclude that well designed neural networks, i.e. a recurrent LSTM neural network initialized with best current evidence and balanced incentives – can potentially advance the protection offered to institutional investors by VaR thresholds through a reduction in threshold breaches. However, such advancements rely on the availability of a long data history, which may not always be available in practice when estimating asset management VaR thresholds. Future work can include the investigation of other asset classes and may also investigate the question if the presented approach is applicable to Value at Risk estimation for banking. From a technical perspective, future research may want to investigate which attributes lead to the performance reduction when halving the input data. Alternatively, future research may attempt to design procedures to make up for a shortage of input data. And, of course, future research may want to test if neural networks could utilize their risk management capabilities ahead of the Covid-19 virus crisis.

# References

Agarwal, V. and Naik, N. Y. 2004. Risks and Portfolio Decisions Involving Hedge Funds. Review of Financial Studies, 17, 1: 63-98.

Albuquerque, R. 2012. Skewness in Stock Returns: Reconciling the Evidence on Firm Versus Aggregate Returns. Review of Financial Studies, 25, 5: 1630–1673.

Aminia, S., Elmore, R., Oztekin, O. and Strauss, J. 2020. Can Machines Learn Capital Structure Dynamics? Working Paper.

Ang, A. and Bekaert, G. 2002. International asset allocation with regime shifts. *The Review of Financial Studies*, 15, 4:1137–1187.

Aubry, M., Kräussl, R., Manso, G., and Spaenjers, C.. 2019. Machine Learning, Human Experts, and the Valuation of Real Assets. Working Paper.

Avramov, D., Cheng, S. and Metzker, L. 2020 Machine Learning versus Economic Restrictions: Evidence from Stock Return Predictability. Working Paper.

Baum, L. E. 1970. A maximization technique occurring in the statistical analysis of probabilistic functions of markov chains, Ann. Math. Statist., 41: 164-171.

Bianchi, D., M. Büchner, and A. Tamoni. 2020. Bond risk premia with machine learning. Working Paper.

Billio, M. and Pelizzon,L.,2000. Value-at-risk: a multivariate switching regime approach. Journal of Empirical Finance 7: 531–554.

Billio, M., Getmansky M., Lo, A. W. and Pelizzon,L.. 2012. Econometric measures of connectedness and systemic risk in the finance and insurance sectors. Journal of Financial Economics. 104: 535-559.

Bishop, C. M. 1995. *Neural networks for pattern recognition*. Oxford university press.

Chen, L., M. Pelger, and J. Zhu. 2020. Deep learning in asset pricing. Working Paper.

De Nard, G., Hedinger, S. and Leippold, M. 2020. Subsampled Factor Models for Asset Pricing: The Rise of Vasa. Working Paper.

Easley, D., López de Prado, M., O'Hara, M. and Zhang, Z.. 2019. Microstructure in the Machine Age. Working Paper.

Fallahgouly, H. and Franstiantoz, V. 2020. Towards Explaining Deep Learning: Significance Tests for Multi-Layer Perceptrons. Working Paper.

Giglio, S. W., and D. Xiu. 2019. Asset Pricing with Omitted Factors. Working Paper.

Götze, T., Gürtler, M. and Witowski, E. 2020. How to Deal with Small Data Sets in Machine Learning: An Analysis on the CAT Bond Market. Working Paper

Glorot, X. and Bengio, Y.. 2010. Understanding the difficulty of training deep feedforward neural networks. Proceedings of the thirteenth international conference on artificial intelligence and statistics. 249–256.

Graves, A. 2013. Generating sequences with recurrent neural networks. *arXiv preprint arXiv:1308.0850*,

Gu, S., B. Kelly, and D. Xiu. 2020. Empirical asset pricing via machine learning. Forthcoming in *Review of Financial Studies*.

Gu, S., B. Kelly, and D. Xiu. 2019. Autoencoder asset pricing models. Forthcoming in *Journal of Econometrics*.

Hann, T.H., Steurer, E., 1996. Much ado about nothing? Exchange rate forecasting: Neural networks vs. linear models using monthly and weekly data. Neurocomputing 10: 323–339

Hochreiter, S. and Schmidhuber, J. 1997. J. Long short-term memory. *Neural computation*, 9, 8:1735–1780..

Horel, E. and Giesecke, K. 2019. Towards explainable ai: Significance tests for neural networks. Working Paper.

Hunt, J.O.S., Myers, J.N., Myers, L.A. 2019. Improving Earnings Predictions with Machine Learning. Working Paper

Jorion, P.. *Value at risk*. 2006, McGraw-Hill Education.

Kelly, B., S. Pruitt, and Y. Su. 2019. Characteristics are covariances: A unified model of risk and return.

Journal of Financial Economics 134:501–524.

Kingma, D. P. and Ba, J. 2014. Adam: A method for stochastic optimization. *arXiv preprint arXiv:1412.6980*

Kozak, S., S. Nagel, and S. Santosh. 2020. Shrinking the cross-section. Journal of Financial Economics 135:271–292.

Kozhan, R., Neuberger, A., Schneider P. 2013. The Skew Risk Premium in the Equity Index Market. Review of Financial Studies, 26, 9: 2174–2203.

Krogh, A. and Hertz, J. A. 1992. A simple weight decay can improve generalization. In *Advances in neural information processing systems*, 950–957.

López de Prado, M. 2018. Advances in Financial Machine Learning. Wiley: Hoboken.

Marcus, G. 2018. Deep Learning: a critical appraisal. *arXiv preprint* arXiv:1801.00631

Nieto, M.R and Ruiz, E. 2015. Frontiers in VaR forecasting and backtesting. International Journal of Forecasting. 32, 2: 475–501.

Patton, A. J., Ziegel, J. F., & Chen, R. 2019. Dynamic semiparametric models for expected shortfall (and value-at-risk). Journal of Econometrics, 211(2), 388–413.

Schmeding, C., Klawunn, M., and Weisheit, S., (2019). Maschinelles Lernen bei der Entwicklung von Wertsicherungsstrategien. Zeitschrift für das gesamte Kreditwesen, 2-2019.

Sirignano, J., Sadhwani, A., Giesecke, K. 2018. Deep learning for mortgage risk. Working paper

Taylor, J. W. 2019. Forecasting value at risk and expected shortfall using a semiparametric approach based on the asymmetric Laplace distribution. Journal of Business & Economic Statistics, 37, 121–133.

Taylor, J. W. 2020. Forecast combinations for value at risk and expected shortfall. International Journal of Forecasting. 36: 428-441.

Timmermann, A. 2000. Moments of markov switching models. *Journal of Econometrics*, 96, 1: 75–111.

van den Oord, A., Dieleman, S., Zen, H., Simonyan, K., Vinyals, O., Graves, A., Kalchbrenner, N., Senior, A. and Koray Kavukcuoglu, K. 2016. Wavenet: A generative model for raw audio. *arXiv preprint arXiv:1609.03499*,

Verstyuk, S. 2020. Modeling Multivariate Time Series in Economics: From Auto-Regressions to Recurrent Neural Networks. Working Paper.

Zhang, H., Dauphin Y. N., Ma T. 2019. Fixup initialization: Residual learning without normalization. Interantional Conference on Learning Representations (ICLR) Paper.

Table 1, Panel A: Descriptive statistics of test assets on daily basis

| Region | US | | | EU | | | UK | | | Global | | |
|---|---|---|---|---|---|---|---|---|---|---|---|---|
| Type | Equity | SB1-3y | SB7-10y | Equity | SB1-3y | SB7-10y | Equity | SB1-3y | SB7-10y | Equity | SB1-3y | SB7-10y |
| Start Date | 5/1/88 | 5/1/88 | 5/1/88 | 2/1/90 | 2/1/90 | 2/1/90 | 2/1/89 | 2/1/89 | 2/1/89 | 2/1/87 | 2/1/87 | 2/1/87 |
| End Date | 31/8/18 | 31/8/18 | 31/8/18 | 31/8/18 | 31/8/18 | 31/8/18 | 31/8/18 | 31/8/18 | 31/8/18 | 31/8/18 | 31/8/18 | 31/8/18 |
| Observations | 7999 | 7999 | 7999 | 7479 | 7479 | 7479 | 7740 | 7740 | 7740 | 8148 | 8148 | 8148 |
| Mean | 0.0004 | 0.0002 | 0.0003 | 0.0003 | 0.0002 | 0.0002 | 0.0004 | 0.0002 | 0.0003 | 0.0004 | 0.0002 | 0.0002 |
| St. Deviation | 0.0108 | 0.0009 | 0.0038 | 0.0132 | 0.0007 | 0.0029 | 0.0107 | 0.0010 | 0.0034 | 0.0092 | 0.0009 | 0.0039 |
| Skewness | -0.1356 | 0.1708 | -0.0908 | 0.0277 | 0.0015 | -0.4182 | -0.0043 | 2.0154 | 0.1509 | -0.3875 | 0.3035 | 0.0142 |
| Minimum | -0.0903 | -0.0088 | -0.0243 | -0.0862 | -0.0061 | -0.0226 | -0.0885 | -0.0067 | -0.0196 | -0.0984 | -0.0088 | -0.0243 |
| 1% | -0.0297 | -0.0022 | -0.0102 | -0.0380 | -0.0018 | -0.0081 | -0.0298 | -0.0026 | -0.0089 | -0.0260 | -0.0022 | -0.0103 |
| 5% | -0.0164 | -0.0011 | -0.0059 | -0.0207 | -0.0010 | -0.0045 | -0.0158 | -0.0013 | -0.0054 | -0.0137 | -0.0012 | -0.0060 |
| 10% | -0.0109 | -0.0008 | -0.0043 | -0.0141 | -0.0006 | -0.0031 | -0.0112 | -0.0008 | -0.0038 | -0.0093 | -0.0008 | -0.0044 |
| 25% | -0.0040 | -0.0003 | -0.0019 | -0.0057 | -0.0002 | -0.0013 | -0.0049 | -0.0003 | -0.0015 | -0.0038 | -0.0003 | -0.0019 |
| 50% | 0.0004 | 0.0001 | 0.0002 | 0.0005 | 0.0001 | 0.0003 | 0.0003 | 0.0002 | 0.0003 | 0.0006 | 0.0001 | 0.0003 |
| 75% | 0.0054 | 0.0006 | 0.0025 | 0.0066 | 0.0005 | 0.0019 | 0.0059 | 0.0007 | 0.0022 | 0.0048 | 0.0006 | 0.0025 |
| 90% | 0.0114 | 0.0011 | 0.0047 | 0.0140 | 0.0010 | 0.0036 | 0.0116 | 0.0012 | 0.0042 | 0.0096 | 0.0012 | 0.0048 |
| 95% | 0.0160 | 0.0016 | 0.0062 | 0.0202 | 0.0013 | 0.0047 | 0.0163 | 0.0017 | 0.0056 | 0.0134 | 0.0016 | 0.0062 |
| 99% | 0.0294 | 0.0027 | 0.0101 | 0.0354 | 0.0021 | 0.0073 | 0.0298 | 0.0030 | 0.0086 | 0.0239 | 0.0027 | 0.0103 |
| Maximum | 0.1158 | 0.0080 | 0.0352 | 0.1100 | 0.0065 | 0.0181 | 0.0984 | 0.0229 | 0.0311 | 0.0952 | 0.0082 | 0.0364 |

Notes: This table displays the descriptive statistics of the daily returns of the main equity index (Equity), the main sovereign bond with (short) 1-3 years maturity (SB1-3y) and the main sovereign bond (long) with 7-10 year maturity (SB7-10). Descriptive statistics include sample length, the first three moments of the return distribution and 11 thresholds along the return distribution.

Table 1, Panel B: Descriptive statistics of test assets on weekly basis

| Region | US | | | EU | | | UK | | | Global | | |
|---|---|---|---|---|---|---|---|---|---|---|---|---|
| Type | Equity | SB1-3y | SB7-10y | Equity | SB1-3y | SB7-10y | Equity | SB1-3y | SB7-10y | Equity | SB1-3y | SB7-10y |
| Start Date | 15/1/88 | 15/1/88 | 15/1/88 | 12/1/90 | 12/1/90 | 12/1/90 | 6/1/89 | 6/1/89 | 6/1/89 | 9/1/87 | 9/1/87 | 9/1/87 |
| End Date | 31/8/18 | 31/8/18 | 31/8/18 | 31/8/18 | 31/8/18 | 31/8/18 | 31/8/18 | 31/8/18 | 31/8/18 | 31/8/18 | 31/8/18 | 31/8/18 |
| Observations | 1599 | 1599 | 1599 | 1495 | 1495 | 1495 | 1548 | 1548 | 1548 | 1652 | 1652 | 1652 |
| Mean | 0.0022 | 0.0008 | 0.0013 | 0.0017 | 0.0008 | 0.0012 | 0.0019 | 0.0010 | 0.0014 | 0.0018 | 0.0008 | 0.0012 |
| St. Deviation | 0.0222 | 0.0020 | 0.0085 | 0.0278 | 0.0018 | 0.0068 | 0.0227 | 0.0024 | 0.0080 | 0.0214 | 0.0021 | 0.0087 |
| Skewness | -0.5050 | 0.2231 | -0.3444 | -0.4294 | 0.7097 | -0.4098 | -0.4265 | 1.0433 | 0.0559 | -0.8283 | 0.7686 | -0.0231 |
| Minimum | -0.1814 | -0.0099 | -0.0380 | -0.2219 | -0.0059 | -0.0373 | -0.2101 | -0.0096 | -0.0325 | -0.2002 | -0.0099 | -0.0380 |
| 1% | -0.0612 | -0.0043 | -0.0223 | -0.0699 | -0.0039 | -0.0165 | -0.0581 | -0.0050 | -0.0191 | -0.0563 | -0.0043 | -0.0223 |
| 5% | -0.0337 | -0.0022 | -0.0131 | -0.0442 | -0.0019 | -0.0107 | -0.0321 | -0.0025 | -0.0118 | -0.0314 | -0.0022 | -0.0135 |
| 10% | -0.0225 | -0.0014 | -0.0094 | -0.0322 | -0.0011 | -0.0075 | -0.0239 | -0.0014 | -0.0084 | -0.0222 | -0.0014 | -0.0095 |
| 25% | -0.0095 | -0.0003 | -0.0039 | -0.0136 | -0.0002 | -0.0026 | -0.0109 | -0.0002 | -0.0032 | -0.0093 | -0.0003 | -0.0039 |
| 50% | 0.0032 | 0.0006 | 0.0014 | 0.0033 | 0.0006 | 0.0017 | 0.0030 | 0.0009 | 0.0016 | 0.0028 | 0.0006 | 0.0013 |
| 75% | 0.0142 | 0.0019 | 0.0071 | 0.0176 | 0.0017 | 0.0055 | 0.0143 | 0.0022 | 0.0062 | 0.0132 | 0.0019 | 0.0070 |
| 90% | 0.0263 | 0.0033 | 0.0114 | 0.0324 | 0.0029 | 0.0091 | 0.0263 | 0.0037 | 0.0105 | 0.0247 | 0.0033 | 0.0114 |
| 95% | 0.0359 | 0.0043 | 0.0140 | 0.0423 | 0.0037 | 0.0116 | 0.0355 | 0.0046 | 0.0132 | 0.0328 | 0.0044 | 0.0140 |
| 99% | 0.0576 | 0.0061 | 0.0201 | 0.0721 | 0.0057 | 0.0167 | 0.0610 | 0.0080 | 0.0220 | 0.0535 | 0.0060 | 0.0202 |
| Maximum | 0.1209 | 0.0106 | 0.0364 | 0.1456 | 0.0161 | 0.0265 | 0.1345 | 0.0216 | 0.0446 | 0.1241 | 0.0218 | 0.0708 |

Notes: This table displays the descriptive statistics of the weekly returns of the main equity index (Equity), the main sovereign bond with (short) 1-3 years maturity (SB1-3y) and the main sovereign bond (long) with 7-10 year maturity (SB7-10). Descriptive statistics include sample length, the first three moments of the return distribution and 11 thresholds along the return distribution.

Table 2, Panel A: 1% VaR thresholds for equity compared with and without initialization

| region | US total: 1197 | | EU total: 1093 | | UK total: 1146 | | GL total: 1250 | |
|---|---|---|---|---|---|---|---|---|
| | breaches | perc | breaches | perc | breaches | perc | breaches | perc |
| model | | | | | | | | |
| Classic | 26 | 2.17% | 27 | 2.47% | 28 | 2.44% | 39 | 3.12% |
| HMM | 16 | 1.34% | 21 | 1.92% | 19 | 1.66% | 27 | 2.16% |
| FF(no_hmm) | 31 | 2.59% | 26 | 2.38% | 32 | 2.79% | 42 | 3.36% |
| CNN(no_hmm) | 28 | 2.34% | 31 | 2.84% | 30 | 2.62% | 48 | 3.84% |
| LSTM(no_hmm) | 480 | 40.10% | 438 | 40.07% | 477 | 41.62% | 511 | 40.88% |
| FF(hmm_init) | 21 | 1.75% | 26 | 2.38% | 26 | 2.27% | 31 | 2.48% |
| CNN(hmm_init) | 22 | 1.84% | 27 | 2.47% | 26 | 2.27% | 35 | 2.80% |
| LSTM(hmm_init) | 19 | 1.59% | 25 | 2.29% | 22 | 1.92% | 33 | 2.64% |

Notes: This table displays the number of VaR breaches (absolute and relative) that each model exhibits. The top rows denote the region of backtesting and the number of total observations. The columns identify the respective models

Table 2, Panel B: 1% VaR thresholds for long bonds compared with and without initialization

| region | US total: 1197 | | EU total: 1093 | | UK total: 1146 | | GL total: 1250 | |
|---|---|---|---|---|---|---|---|---|
| model | breaches | perc | breaches | perc | breaches | perc | breaches | perc |
| Classic | 27 | 2.26% | 27 | 2.47% | 22 | 1.92% | 26 | 2.08% |
| HMM | 21 | 1.75% | 21 | 1.92% | 16 | 1.40% | 20 | 1.60% |
| FF(no_hmm) | 37 | 3.09% | 40 | 3.66% | 35 | 3.05% | 33 | 2.64% |
| CNN(no_hmm) | 45 | 3.76% | 38 | 3.48% | 27 | 2.36% | 50 | 4.00% |
| LSTM(no_hmm) | 480 | 40.10% | 384 | 35.13% | 447 | 39.01% | 493 | 39.44% |
| FF(hmm_init) | 28 | 2.34% | 28 | 2.56% | 27 | 2.36% | 35 | 2.80% |
| CNN(hmm_init) | 30 | 2.51% | 35 | 3.20% | 26 | 2.27% | 32 | 2.56% |
| LSTM(hmm_init) | 24 | 2.01% | 29 | 2.65% | 20 | 1.75% | 26 | 2.08% |

Notes: This table displays the number of VaR breaches (absolute and relative) that each model exhibits. The top rows denote the region of backtesting and the number of total observations. The columns identify the respective models

Table 2, Panel C: 5% VaR thresholds for equity compared with and without initialization

| region<br><br>model | US<br>total: 1197 | | EU<br>total: 1093 | | UK<br>total: 1146 | | GL<br>total: 1250 | |
|---|---|---|---|---|---|---|---|---|
| | breaches | perc | breaches | perc | breaches | perc | breaches | perc |
| Classic | 64 | 5.35% | 69 | 6.31% | 64 | 5.58% | 84 | 6.72% |
| HMM | 58 | 4.85% | 68 | 6.22% | 62 | 5.41% | 74 | 5.92% |
| FF(no_hmm) | 67 | 5.60% | 70 | 6.40% | 67 | 5.85% | 90 | 7.20% |
| CNN(no_hmm) | 75 | 6.27% | 77 | 7.04% | 68 | 5.93% | 103 | 8.24% |
| LSTM(no_hmm) | 501 | 41.85% | 461 | 42.18% | 490 | 42.76% | 526 | 42.08% |
| FF(hmm_init) | 60 | 5.01% | 72 | 6.59% | 67 | 5.85% | 87 | 6.96% |
| CNN(hmm_init) | 77 | 6.43% | 84 | 7.69% | 71 | 6.20% | 111 | 8.88% |
| LSTM(hmm_init) | 61 | 5.10% | 69 | 6.31% | 62 | 5.41% | 80 | 6.40% |

Notes: This table displays the number of VaR breaches (absolute and relative) that each model exhibits. The top rows denote the region of backtesting and the number of total observations. The columns identify the respective models

Table 2, Panel D: 5% VaR thresholds for long bonds compared with and without initialization

| model | US total: 1197 | | EU total: 1093 | | UK total: 1146 | | GL total: 1250 | |
|---|---|---|---|---|---|---|---|---|
| region | breaches | perc | breaches | perc | breaches | perc | breaches | perc |
| Classic | 56 | 4.68% | 68 | 6.22% | 50 | 4.36% | 56 | 4.48% |
| HMM | 60 | 5.01% | 72 | 6.59% | 56 | 4.89% | 60 | 4.80% |
| FF(no_hmm) | 99 | 8.27% | 94 | 8.60% | 79 | 6.89% | 87 | 6.96% |
| CNN(no_hmm) | 110 | 9.19% | 90 | 8.23% | 79 | 6.89% | 112 | 8.96% |
| LSTM(no_hmm) | 501 | 41.85% | 420 | 38.43% | 484 | 42.23% | 515 | 41.20% |
| FF(hmm_init) | 70 | 5.85% | 71 | 6.50% | 60 | 5.24% | 77 | 6.16% |
| CNN(hmm_init) | 79 | 6.60% | 86 | 7.87% | 82 | 7.16% | 91 | 7.28% |
| LSTM(hmm_init) | 65 | 5.43% | 76 | 6.95% | 55 | 4.80% | 65 | 5.20% |

Notes: This table displays the number of VaR breaches (absolute and relative) that each model exhibits. The top rows denote the region of backtesting and the number of total observations. The columns identify the respective models

Table 3, Panel A: 1% VaR thresholds for equity compared with and without balanced incentive function (i.e. regularized)

| region | US total: 1197 | | EU total: 1093 | | UK total: 1146 | | GL total: 1250 | |
|---|---|---|---|---|---|---|---|---|
| model | breaches | perc | breaches | perc | breaches | perc | breaches | perc |
| Classic | 26 | 2.17% | 26 | 2.38% | 28 | 2.44% | 39 | 3.12% |
| HMM | 16 | 1.34% | 21 | 1.92% | 19 | 1.66% | 27 | 2.16% |
| FF (hmm init) | 21 | 1.75% | 26 | 2.38% | 26 | 2.27% | 31 | 2.48% |
| CNN (hmm init) | 22 | 1.84% | 27 | 2.47% | 26 | 2.27% | 35 | 2.80% |
| LSTM (hmm init) | 19 | 1.59% | 25 | 2.29% | 22 | 1.92% | 33 | 2.64% |
| FF (hmm init + reg) | 22 | 1.84% | 24 | 2.20% | 25 | 2.18% | 30 | 2.40% |
| CNN (hmm init + reg) | 26 | 2.17% | 29 | 2.65% | 37 | 3.23% | 49 | 3.92% |
| LSTM (hmm init + reg) | 19 | 1.59% | 21 | 1.92% | 21 | 1.83% | 27 | 2.16% |

Notes: This table displays the number of VaR breaches (absolute and relative) that each model exhibits. The top rows denote the region of backtesting and the number of total observations. The columns identify the respective models

Table 3, Panel B: 1% VaR thresholds for long bonds compared with and without balanced incentive function (i.e. regularized)

| model | US total: 1197 breaches | perc | EU total: 1093 breaches | perc | UK total: 1146 breaches | perc | GL total: 1250 breaches | perc |
|---|---|---|---|---|---|---|---|---|
| Classic | 26 | 2.17% | 27 | 2.47% | 23 | 2.01% | 27 | 2.16% |
| HMM | 21 | 1.75% | 21 | 1.92% | 16 | 1.40% | 20 | 1.60% |
| FF (hmm init) | 28 | 2.34% | 28 | 2.56% | 27 | 2.36% | 35 | 2.80% |
| CNN (hmm init) | 30 | 2.51% | 35 | 3.20% | 26 | 2.27% | 32 | 2.56% |
| LSTM (hmm init) | 24 | 2.01% | 29 | 2.65% | 20 | 1.75% | 26 | 2.08% |
| FF (hmm init + reg) | 26 | 2.17% | 27 | 2.47% | 23 | 2.01% | 30 | 2.40% |
| CNN (hmm init + reg) | 30 | 2.51% | 36 | 3.29% | 25 | 2.18% | 34 | 2.72% |
| LSTM (hmm init + reg) | 16 | 1.34% | 21 | 1.92% | 14 | 1.22% | 19 | 1.52% |

Notes: This table displays the number of VaR breaches (absolute and relative) that each model exhibits. The top rows denote the region of backtesting and the number of total observations. The columns identify the respective models

Table 3, Panel C: 5% VaR thresholds for equity compared with and without balanced incentive function (i.e. regularized)

| region | US total: 1197 | | EU total: 1093 | | UK total: 1146 | | GL total: 1250 | |
|---|---|---|---|---|---|---|---|---|
| | breaches | perc | breaches | perc | breaches | perc | breaches | perc |
| model | | | | | | | | |
| Classic | 65 | 5.43% | 69 | 6.31% | 65 | 5.67% | 82 | 6.56% |
| HMM | 58 | 4.85% | 68 | 6.22% | 62 | 5.41% | 74 | 5.92% |
| FF (hmm init) | 60 | 5.01% | 72 | 6.59% | 67 | 5.85% | 87 | 6.96% |
| CNN (hmm init) | 77 | 6.43% | 84 | 7.69% | 71 | 6.20% | 111 | 8.88% |
| LSTM (hmm init) | 61 | 5.10% | 69 | 6.31% | 62 | 5.41% | 80 | 6.40% |
| FF (hmm init + reg) | 60 | 5.01% | 67 | 6.13% | 60 | 5.24% | 79 | 6.32% |
| CNN (hmm init + reg) | 70 | 5.85% | 83 | 7.59% | 79 | 6.89% | 99 | 7.92% |
| LSTM (hmm init + reg) | 46 | 3.84% | 55 | 5.03% | 52 | 4.54% | 64 | 5.12% |

Notes: This table displays the number of VaR breaches (absolute and relative) that each model exhibits. The top rows denote the region of backtesting and the number of total observations. The columns identify the respective models

Table 3, Panel D: 5% VaR thresholds for long bonds compared with and without balanced incentive function (i.e. regularized)

| region | US total: 1197 | | EU total: 1093 | | UK total: 1146 | | GL total: 1250 | |
|---|---|---|---|---|---|---|---|---|
| | breaches | perc | breaches | perc | breaches | perc | breaches | perc |
| **model** | | | | | | | | |
| **Classic** | 58 | 4.85% | 68 | 6.22% | 50 | 4.36% | 57 | 4.56% |
| **HMM** | 60 | 5.01% | 72 | 6.59% | 56 | 4.89% | 60 | 4.80% |
| **FF (hmm init)** | 70 | 5.85% | 71 | 6.50% | 60 | 5.24% | 77 | 6.16% |
| **CNN (hmm init)** | 79 | 6.60% | 86 | 7.87% | 82 | 7.16% | 91 | 7.28% |
| **LSTM (hmm init)** | 65 | 5.43% | 76 | 6.95% | 55 | 4.80% | 65 | 5.20% |
| **FF (hmm init + reg)** | 63 | 5.26% | 69 | 6.31% | 52 | 4.54% | 68 | 5.44% |
| **CNN (hmm init + reg)** | 70 | 5.85% | 93 | 8.51% | 75 | 6.54% | 79 | 6.32% |
| **LSTM (hmm init + reg)** | 42 | 3.51% | 59 | 5.40% | 39 | 3.40% | 48 | 3.84% |

Notes: This table displays the number of VaR breaches (absolute and relative) that each model exhibits. The top rows denote the region of backtesting and the number of total observations. The columns identify the respective models

**Table 4, Panel A:**   Comparison of estimated 1% VaR of equities with and without balanced incentive function (i.e. regularized)

| model | region | Classic | | | HMM | | | FF (hmm init) | | | CNN (hmm init) | | | LSTM (hmm init) | | | FF (hmm init + reg) | | | CNN (hmm init + reg) | | | LSTM (hmm init + reg) | | |
|---|---|---|---|---|---|---|---|---|---|---|---|---|---|---|---|---|---|---|---|---|---|---|---|---|---|---|
| | | comp | pvalue | dom | comp | pvalue | dom | comp | pvalue | dom | comp | pvalue | dom | comp | pvalue | dom | comp | pvalue | dom | comp | pvalue | dom | comp | pvalue | dom |
| **Classic** | US | | | | 0.0 | 0.0009 | -0.41 | 0.0 | 0.0577 | -0.30 | 0.0 | 0.1656 | -0.33 | 0.0 | 0.0209 | -0.37 | 0.0 | 0.0588 | -0.22 | 0.0 | 0.7964 | -0.30 | 0.0 | 0.0114 | -0.33 |
| | EU | | | | 0.0 | 0.0577 | -0.30 | 0.0 | 0.7390 | -0.19 | 0.5 | 1.0000 | -0.26 | 0.0 | 0.5932 | -0.30 | 0.0 | 0.1798 | -0.15 | 1.0 | 0.5639 | -0.19 | 0.0 | 0.0338 | -0.26 |
| | UK | | | | 0.0 | 0.0046 | -0.30 | 0.0 | 0.7632 | -0.22 | 1.0 | 0.7964 | -0.30 | 0.0 | 0.0956 | -0.26 | 0.0 | 0.4145 | -0.15 | 1.0 | 0.0329 | -0.22 | 0.0 | 0.0338 | -0.26 |
| | GL | | | | 0.0 | 0.0005 | -0.31 | 0.0 | 0.0046 | -0.21 | 0.0 | 0.4144 | -0.34 | 0.0 | 0.0833 | -0.23 | 0.0 | 0.0027 | -0.23 | 0.0 | 0.0588 | -0.23 | 0.0 | 0.0013 | -0.33 |
| **HMM** | US | 1.0 | 0.0009 | 0.00 | | | | 1.0 | 0.0956 | -0.12 | 1.0 | 0.0577 | -0.12 | 1.0 | 0.1798 | -0.06 | 1.0 | 0.0338 | -0.06 | 1.0 | 0.0075 | -0.12 | 1.0 | 0.1798 | -0.06 |
| | EU | 1.0 | 0.0577 | -0.10 | | | | 1.0 | 0.0956 | -0.10 | 1.0 | 0.0833 | -0.14 | 1.0 | 0.0455 | 0.00 | 1.0 | 0.2570 | -0.10 | 1.0 | 0.0324 | -0.14 | 0.5 | 1.0000 | -0.10 |
| | UK | 1.0 | 0.0046 | 0.00 | | | | 1.0 | 0.0196 | -0.05 | 1.0 | 0.0522 | -0.16 | 1.0 | 0.1798 | -0.05 | 1.0 | 0.0338 | -0.05 | 1.0 | 0.0001 | -0.11 | 1.0 | 0.4797 | -0.16 |
| | GL | 1.0 | 0.0005 | 0.00 | | | | 1.0 | 0.1574 | -0.07 | 1.0 | 0.0593 | -0.19 | 1.0 | 0.0338 | -0.04 | 1.0 | 0.2570 | -0.07 | 1.0 | 0.0000 | -0.04 | 0.5 | 1.0000 | -0.11 |
| **FF (hmm init)** | US | 1.0 | 0.0577 | -0.10 | 0.0 | 0.0956 | -0.33 | | | | 1.0 | 0.7964 | -0.33 | 1.0 | 0.5273 | -0.29 | 1.0 | 0.6549 | -0.10 | 1.0 | 0.2254 | -0.29 | 1.0 | 0.5273 | -0.29 |
| | EU | 1.0 | 0.7390 | -0.15 | 0.0 | 0.0956 | -0.27 | | | | 1.0 | 0.7964 | -0.27 | 1.0 | 0.7632 | -0.23 | 1.0 | 0.4145 | -0.15 | 1.0 | 0.4056 | -0.19 | 0.0 | 0.1317 | -0.31 |
| | UK | 1.0 | 0.7632 | -0.19 | 0.0 | 0.0196 | -0.31 | | | | 0.5 | 1.0000 | -0.31 | 1.0 | 0.2060 | -0.27 | 1.0 | 0.7056 | -0.15 | 1.0 | 0.0116 | -0.35 | 0.0 | 0.1656 | -0.35 |
| | GL | 1.0 | 0.0046 | 0.00 | 0.0 | 0.1574 | -0.19 | | | | 1.0 | 0.3460 | -0.23 | 1.0 | 0.4797 | -0.10 | 1.0 | 0.5639 | -0.06 | 1.0 | 0.0002 | -0.10 | 1.0 | 0.1025 | -0.16 |
| **CNN (hmm init)** | US | 1.0 | 0.1656 | -0.18 | 0.0 | 0.0577 | -0.36 | 0.0 | 0.7964 | -0.36 | | | | 0.0 | 0.4388 | -0.41 | 0.5 | 1.0000 | -0.36 | 1.0 | 0.2484 | -0.18 | 1.0 | 0.4056 | -0.36 |
| | EU | 0.5 | 1.0000 | -0.24 | 0.0 | 0.0833 | -0.33 | 0.0 | 0.7964 | -0.24 | | | | 0.0 | 0.5932 | -0.30 | 0.0 | 0.4056 | -0.30 | 1.0 | 0.6376 | -0.24 | 1.0 | 0.1088 | -0.37 |
| | UK | 1.0 | 0.7964 | -0.27 | 0.0 | 0.0522 | -0.38 | 0.5 | 1.0000 | -0.31 | | | | 0.0 | 0.2852 | -0.35 | 0.0 | 0.8085 | -0.35 | 1.0 | 0.0277 | -0.27 | 1.0 | 0.2254 | -0.42 |
| | GL | 1.0 | 0.4144 | -0.29 | 0.0 | 0.0593 | -0.37 | 1.0 | 0.3460 | -0.31 | | | | 0.0 | 0.6549 | -0.31 | 0.0 | 0.2254 | -0.31 | 1.0 | 0.0028 | -0.11 | 0.0 | 0.0736 | -0.40 |
| **LSTM (hmm init)** | US | 1.0 | 0.0209 | -0.11 | 1.0 | 0.1798 | -0.21 | 1.0 | 0.5273 | -0.21 | 1.0 | 0.4388 | -0.32 | | | | 1.0 | 0.3175 | -0.16 | 1.0 | 0.0707 | -0.12 | 0.5 | 1.0000 | -0.11 |
| | EU | 1.0 | 0.5932 | -0.24 | 1.0 | 0.0455 | -0.16 | 1.0 | 0.7632 | -0.20 | 1.0 | 0.5932 | -0.24 | | | | 1.0 | 0.7632 | -0.20 | 1.0 | 0.3460 | -0.28 | 1.0 | 0.1574 | -0.24 |
| | UK | 1.0 | 0.0956 | -0.09 | 1.0 | 0.1798 | -0.18 | 1.0 | 0.2060 | -0.14 | 1.0 | 0.2852 | -0.23 | | | | 1.0 | 0.2570 | -0.09 | 1.0 | 0.0010 | -0.14 | 1.0 | 0.7390 | -0.23 |
| | GL | 1.0 | 0.0833 | -0.09 | 1.0 | 0.0338 | -0.21 | 1.0 | 0.4797 | -0.15 | 1.0 | 0.6549 | -0.27 | | | | 0.0 | 0.2570 | -0.15 | 1.0 | 0.0011 | -0.12 | 0.0 | 0.0338 | -0.21 |
| **FF (hmm init + reg)** | US | 1.0 | 0.0588 | -0.05 | 0.0 | 0.0338 | -0.32 | 0.0 | 0.6549 | -0.14 | 0.0 | 1.0000 | -0.36 | 0.0 | 0.3175 | -0.27 | | | | 1.0 | 0.3175 | -0.27 | 1.0 | 0.2570 | -0.23 |
| | EU | 1.0 | 0.1798 | -0.04 | 1.0 | 0.2570 | -0.21 | 1.0 | 0.4145 | -0.08 | 1.0 | 0.4056 | -0.21 | 0.0 | 0.7632 | -0.08 | | | | 1.0 | 0.1317 | -0.12 | 1.0 | 0.2570 | -0.21 |
| | UK | 1.0 | 0.4145 | -0.04 | 1.0 | 0.0338 | -0.28 | 0.0 | 0.7056 | -0.15 | 1.0 | 0.8085 | -0.20 | 0.0 | 0.2570 | -0.09 | | | | 1.0 | 0.0105 | -0.20 | 1.0 | 0.2060 | -0.28 |
| | GL | 1.0 | 0.0027 | 0.00 | 1.0 | 0.2570 | -0.17 | 0.0 | 0.5639 | -0.03 | 1.0 | 0.2254 | -0.20 | 1.0 | 0.2570 | -0.07 | | | | 1.0 | 0.0001 | -0.07 | 1.0 | 0.1798 | -0.13 |
| **CNN (hmm init + reg)** | US | 0.0 | 0.7964 | -0.27 | 1.0 | 0.0075 | -0.46 | 1.0 | 0.2254 | -0.42 | 0.0 | 0.2484 | -0.31 | 0.0 | 0.0707 | -0.42 | 0.0 | 0.3175 | -0.38 | | | | 0.0 | 0.0896 | -0.46 |
| | EU | 0.0 | 0.5639 | -0.24 | 1.0 | 0.0324 | -0.38 | 1.0 | 0.4056 | -0.28 | 0.0 | 0.6376 | -0.34 | 0.0 | 0.3460 | -0.28 | 0.0 | 0.1317 | -0.12 | | | | 0.0 | 0.0455 | -0.41 |
| | UK | 1.0 | 0.0329 | -0.43 | 1.0 | 0.0001 | -0.54 | 1.0 | 0.0116 | -0.41 | 0.0 | 0.0277 | -0.49 | 0.0 | 0.0010 | -0.49 | 0.0 | 0.0105 | -0.46 | | | | 0.0 | 0.0006 | -0.51 |
| | GL | 0.0 | 0.0588 | -0.39 | 1.0 | 0.0000 | -0.47 | 1.0 | 0.0002 | -0.43 | 0.0 | 0.0028 | -0.37 | 0.0 | 0.0011 | -0.41 | 0.0 | 0.0001 | -0.43 | | | | 0.0 | 0.0000 | -0.45 |
| **LSTM (hmm init + reg)** | US | 1.0 | 0.0114 | -0.05 | 1.0 | 0.1798 | -0.21 | 1.0 | 0.5273 | -0.21 | 1.0 | 0.4056 | -0.26 | 0.5 | 1.0000 | -0.11 | 1.0 | 0.2570 | -0.11 | 1.0 | 0.0896 | -0.46 | | | |
| | EU | 1.0 | 0.0338 | -0.05 | 0.5 | 1.0000 | -0.10 | 1.0 | 0.1317 | -0.14 | 1.0 | 0.1088 | -0.19 | 1.0 | 0.1574 | -0.10 | 1.0 | 0.2570 | -0.11 | 1.0 | 0.0455 | -0.19 | | | |
| | UK | 1.0 | 0.0338 | -0.05 | 0.0 | 0.4797 | -0.24 | 1.0 | 0.1656 | -0.19 | 1.0 | 0.2254 | -0.29 | 1.0 | 0.7390 | -0.19 | 1.0 | 0.2060 | -0.14 | 1.0 | 0.0006 | -0.14 | | | |
| | GL | 1.0 | 0.0013 | -0.04 | 0.5 | 1.0000 | -0.11 | 1.0 | 0.1025 | -0.16 | 1.0 | 0.0736 | -0.11 | 1.0 | 0.0338 | -0.21 | 1.0 | 0.1798 | -0.13 | 1.0 | 0.0000 | -0.45 | | | |

Notes: This table displays the model comparison as a matrix. The models are listed per row (assessed model) and are compared to their peers which are listed per column (benchmark model). If a value indicate an advantage of one model over the other, it is colored green, otherwise red. P-values are colored if significant (< 0.1). Dominance (dom) values are coloured green is dominant (i.e. 0.00) and red if clearly not dominant (i.e. ≤ -0.1).

**Table 4, Panel B:**  Comparison of estimated 1% VaR of long bonds with and without balanced incentive function (i.e. regularized)

| model | region (peer) | Classic comp | Classic pvalue | Classic dom | HMM comp | HMM pvalue | HMM dom | FF (hmm init) comp | FF (hmm init) pvalue | FF (hmm init) dom | CNN (hmm init) comp | CNN (hmm init) pvalue | CNN (hmm init) dom | LSTM (hmm init) comp | LSTM (hmm init) pvalue | LSTM (hmm init) dom | FF (hmm init + reg) comp | FF (hmm init + reg) pvalue | FF (hmm init + reg) dom | CNN (hmm init + reg) comp | CNN (hmm init + reg) pvalue | CNN (hmm init + reg) dom | LSTM (hmm init + reg) comp | LSTM (hmm init + reg) pvalue | LSTM (hmm init + reg) dom |
|---|---|---|---|---|---|---|---|---|---|---|---|---|---|---|---|---|---|---|---|---|---|---|---|---|---|
| Classic | US | | | | 0.0 | 0.0253 | -0.19 | 1.0 | 0.4144 | -0.08 | 1.0 | 0.2484 | -0.15 | 0.0 | 0.5273 | -0.23 | 0.5 | 1.0000 | -0.12 | 1.0 | 0.1574 | -0.08 | 0.0 | 0.0039 | -0.42 |
| | EU | | | | 0.0 | 0.1968 | -0.38 | 1.0 | 0.4145 | -0.08 | 1.0 | 0.0066 | -0.04 | 1.0 | 0.3175 | -0.12 | 1.0 | 0.6549 | -0.08 | 1.0 | 0.0329 | -0.23 | 0.0 | 0.1317 | -0.31 |
| | UK | | | | 0.0 | 0.0114 | -0.38 | 1.0 | 0.1798 | -0.04 | 1.0 | 0.5273 | -0.17 | 0.0 | 0.1574 | -0.25 | 1.0 | 0.6549 | -0.12 | 1.0 | 0.7964 | -0.29 | 0.0 | 0.0015 | -0.42 |
| | GL | | | | 0.0 | 0.0338 | -0.27 | 1.0 | 0.0027 | 0.00 | 1.0 | 0.0833 | -0.12 | 0.5 | 1.0000 | -0.12 | 1.0 | 0.0455 | 0.00 | 1.0 | 0.0209 | -0.08 | 0.0 | 0.0196 | -0.31 |
| HMM | US | 1.0 | 0.0253 | 0.00 | | | | 1.0 | 0.0081 | 0.00 | 1.0 | 0.0125 | -0.10 | 1.0 | 0.2570 | -0.10 | 1.0 | 0.0253 | 0.00 | 1.0 | 0.0027 | 0.00 | 0.0 | 0.0588 | -0.29 |
| | EU | 1.0 | 0.1968 | -0.24 | | | | 1.0 | 0.0707 | -0.19 | 1.0 | 0.0005 | -0.05 | 1.0 | 0.0209 | -0.10 | 1.0 | 0.1088 | -0.19 | 1.0 | 0.0010 | -0.14 | 0.5 | 1.0000 | -0.19 |
| | UK | 1.0 | 0.0114 | -0.06 | | | | 1.0 | 0.0009 | 0.00 | 1.0 | 0.0039 | -0.06 | 1.0 | 0.0455 | 0.00 | 1.0 | 0.0196 | -0.06 | 1.0 | 0.0066 | -0.06 | 1.0 | 0.3175 | -0.19 |
| | GL | 1.0 | 0.0338 | -0.05 | | | | 1.0 | 0.0046 | -0.15 | 1.0 | 0.0003 | -0.05 | 1.0 | 0.0142 | 0.00 | 1.0 | 0.0039 | -0.05 | 1.0 | 0.0005 | -0.05 | 1.0 | 0.7390 | -0.25 |
| FF (hmm init) | US | 0.0 | 0.4144 | -0.14 | 0.0 | 0.0081 | -0.25 | | | | 1.0 | 0.5639 | -0.18 | 0.0 | 0.2060 | -0.25 | 0.0 | 0.3175 | -0.11 | 0.0 | 0.5273 | -0.14 | 0.0 | 0.0013 | -0.46 |
| | EU | 0.0 | 0.4145 | -0.14 | 0.0 | 0.0707 | -0.39 | | | | 1.0 | 0.0522 | -0.11 | 1.0 | 0.7817 | -0.21 | 0.0 | 0.5639 | -0.07 | 0.0 | 0.0881 | -0.25 | 0.0 | 0.0522 | -0.36 |
| | UK | 0.0 | 0.1798 | -0.15 | 0.0 | 0.0009 | -0.41 | | | | 0.0 | 0.6549 | -0.11 | 1.0 | 0.0196 | -0.04 | 0.0 | 0.1025 | -0.19 | 0.0 | 0.5932 | -0.32 | 0.0 | 0.0003 | -0.48 |
| | GL | 0.0 | 0.0027 | -0.26 | 0.0 | 0.0003 | -0.46 | | | | 1.0 | 0.3175 | -0.17 | 0.0 | 0.0125 | -0.31 | 0.0 | 0.0253 | -0.14 | 0.0 | 0.7632 | -0.17 | 0.0 | 0.0002 | -0.49 |
| CNN (hmm init) | US | 0.0 | 0.2484 | -0.27 | 0.0 | 0.0125 | -0.37 | 0.0 | 0.5639 | -0.23 | | | | 0.0 | 0.0833 | -0.30 | 0.0 | 0.2852 | -0.30 | 0.5 | 1.0000 | -0.23 | 0.0 | 0.0009 | -0.53 |
| | EU | 0.0 | 0.0066 | -0.41 | 0.0 | 0.0005 | -0.43 | 0.0 | 0.0522 | -0.04 | | | | 0.0 | 0.1088 | -0.20 | 0.0 | 0.0324 | -0.31 | 1.0 | 0.8187 | -0.26 | 0.0 | 0.0006 | -0.43 |
| | UK | 0.0 | 0.5273 | -0.23 | 0.0 | 0.0039 | -0.42 | 0.0 | 0.6549 | -0.08 | | | | 0.0 | 0.0577 | -0.31 | 0.0 | 0.3659 | -0.27 | 1.0 | 0.7816 | -0.27 | 0.0 | 0.0027 | -0.54 |
| | GL | 0.0 | 0.0833 | -0.28 | 0.0 | 0.0046 | -0.47 | 1.0 | 0.3175 | -0.09 | | | | 0.0 | 0.1088 | -0.15 | 0.0 | 0.4797 | -0.16 | 1.0 | 0.6173 | -0.22 | 0.0 | 0.0016 | -0.47 |
| LSTM (hmm init) | US | 1.0 | 0.5273 | -0.17 | 0.0 | 0.2570 | -0.21 | 1.0 | 0.2060 | -0.12 | 1.0 | 0.0833 | -0.12 | | | | 1.0 | 0.4797 | -0.12 | 1.0 | 0.0833 | -0.12 | 0.0 | 0.0044 | -0.33 |
| | EU | 1.0 | 0.3175 | -0.21 | 0.0 | 0.0209 | -0.34 | 1.0 | 0.7817 | -0.24 | 1.0 | 0.1088 | -0.14 | | | | 1.0 | 0.5639 | -0.24 | 1.0 | 0.0707 | -0.14 | 0.0 | 0.0113 | -0.31 |
| | UK | 1.0 | 0.1574 | -0.10 | 0.0 | 0.0455 | -0.20 | 1.0 | 0.0196 | -0.05 | 1.0 | 0.0577 | -0.10 | | | | 1.0 | 0.2570 | -0.10 | 1.0 | 0.0956 | -0.10 | 0.0 | 0.0142 | -0.30 |
| | GL | 0.5 | 1.0000 | -0.12 | 0.0 | 0.0142 | -0.23 | 1.0 | 0.0125 | -0.08 | 1.0 | 0.1088 | -0.15 | | | | 1.0 | 0.2060 | -0.12 | 1.0 | 0.0209 | -0.08 | 0.0 | 0.0081 | -0.27 |
| FF (hmm init + reg) | US | 0.5 | 1.0000 | -0.12 | 0.0 | 0.0253 | -0.19 | 1.0 | 0.3175 | -0.04 | 1.0 | 0.2852 | -0.19 | 0.0 | 0.4797 | -0.19 | | | | 1.0 | 0.1574 | -0.08 | 0.0 | 0.0039 | -0.42 |
| | EU | 0.0 | 0.6549 | -0.11 | 0.0 | 0.1088 | -0.37 | 1.0 | 0.5639 | -0.24 | 1.0 | 0.0324 | -0.11 | 1.0 | 0.5639 | -0.19 | | | | 1.0 | 0.0495 | -0.22 | 0.0 | 0.0833 | -0.33 |
| | UK | 1.0 | 0.6549 | -0.09 | 0.0 | 0.0196 | -0.37 | 1.0 | 0.1025 | -0.04 | 1.0 | 0.3659 | -0.17 | 1.0 | 0.2570 | -0.10 | | | | 1.0 | 0.5932 | -0.26 | 0.0 | 0.0027 | -0.39 |
| | GL | 0.0 | 0.0455 | -0.13 | 0.0 | 0.0039 | -0.37 | 1.0 | 0.0253 | 0.00 | 1.0 | 0.4797 | -0.10 | 1.0 | 0.2060 | -0.23 | | | | 1.0 | 0.2484 | -0.13 | 0.0 | 0.0023 | -0.40 |
| CNN (hmm init + reg) | US | 0.0 | 0.1574 | -0.20 | 0.0 | 0.0027 | -0.30 | 0.0 | 0.5273 | -0.20 | 0.5 | 1.0000 | -0.23 | 0.0 | 0.0833 | -0.30 | 0.0 | 0.1574 | -0.20 | | | | 0.0 | 0.0005 | -0.50 |
| | EU | 0.0 | 0.0329 | -0.44 | 0.0 | 0.0010 | -0.50 | 0.0 | 0.0881 | -0.42 | 1.0 | 0.8187 | -0.28 | 0.0 | 0.0707 | -0.31 | 0.0 | 0.0495 | -0.42 | | | | 0.0 | 0.0006 | -0.47 |
| | UK | 0.0 | 0.7964 | -0.32 | 0.0 | 0.0066 | -0.40 | 0.0 | 0.5932 | -0.24 | 1.0 | 0.7816 | -0.24 | 0.0 | 0.0956 | -0.28 | 0.0 | 0.5932 | -0.32 | | | | 0.0 | 0.0045 | -0.52 |
| | GL | 0.0 | 0.0209 | -0.29 | 0.0 | 0.0005 | -0.44 | 0.0 | 0.7632 | -0.15 | 1.0 | 0.6173 | -0.26 | 0.0 | 0.0209 | -0.08 | 0.0 | 0.2484 | -0.24 | | | | 0.0 | 0.0001 | -0.44 |
| LSTM (hmm init + reg) | US | 1.0 | 0.0039 | -0.06 | 1.0 | 0.0588 | -0.06 | 1.0 | 0.0009 | -0.12 | 1.0 | 0.0046 | -0.10 | 1.0 | 0.0039 | -0.06 | 1.0 | 0.0039 | -0.06 | 1.0 | 0.0005 | -0.06 | | | |
| | EU | 1.0 | 0.1317 | -0.14 | 0.5 | 1.0000 | -0.19 | 1.0 | 0.0522 | -0.14 | 1.0 | 0.0005 | -0.05 | 1.0 | 0.0113 | -0.05 | 1.0 | 0.0833 | -0.14 | 1.0 | 0.0006 | -0.10 | | | |
| | UK | 1.0 | 0.0015 | 0.00 | 1.0 | 0.3175 | -0.07 | 1.0 | 0.0003 | -0.14 | 1.0 | 0.0027 | -0.14 | 1.0 | 0.0142 | 0.00 | 1.0 | 0.0027 | 0.00 | 1.0 | 0.0045 | -0.14 | | | |
| | GL | 1.0 | 0.0196 | -0.05 | 1.0 | 0.7390 | -0.21 | 1.0 | 0.0002 | -0.05 | 1.0 | 0.0016 | -0.14 | 1.0 | 0.0081 | 0.00 | 1.0 | 0.0023 | 0.00 | 1.0 | 0.0001 | 0.00 | | | |

Notes: This table displays the model comparison as a matrix. The models are listed per row (assessed model) and are compared to their peers which are listed per column (benchmark model). If a value indicate an advantage of one model over the other, it is colored green, otherwise red. P-values are colored if significant (< 0.1). Dominance (dom) values are coloured green is dominant (i.e. 0.00) and red if clearly not dominant (i.e. ≤ -0.1).

**Table 4, Panel C:** Comparison of estimated 5% VaR of equities with and without balanced incentive function (i.e. regularized)

| model | region | Classic comp | Classic pvalue | Classic dom | HMM comp | HMM pvalue | HMM dom | FF (hmm init) comp | FF (hmm init) pvalue | FF (hmm init) dom | CNN (hmm init) comp | CNN (hmm init) pvalue | CNN (hmm init) dom | LSTM (hmm init) comp | LSTM (hmm init) pvalue | LSTM (hmm init) dom | FF (hmm init + reg) comp | FF (hmm init + reg) pvalue | FF (hmm init + reg) dom | CNN (hmm init + reg) comp | CNN (hmm init + reg) pvalue | CNN (hmm init + reg) dom | LSTM (hmm init + reg) comp | LSTM (hmm init + reg) pvalue | LSTM (hmm init + reg) dom |
|---|---|---|---|---|---|---|---|---|---|---|---|---|---|---|---|---|---|---|---|---|---|---|---|---|---|
| Classic | US | | | | 0.0 | 0.1337 | -0.17 | 0.0 | 0.2852 | -0.14 | 1.0 | 0.0633 | -0.28 | 0.0 | 0.5318 | -0.20 | 0.0 | 0.2060 | -0.11 | 1.0 | 0.3175 | -0.23 | 0.0 | 0.0001 | -0.31 |
| | EU | | | | 0.0 | 0.7056 | -0.21 | 1.0 | 0.5932 | -0.09 | 1.0 | 0.0268 | -0.19 | 0.0 | 0.8619 | -0.24 | 0.0 | 0.4056 | -0.11 | 1.0 | 0.0526 | -0.23 | 0.0 | 0.0038 | -0.30 |
| | UK | | | | 0.0 | 0.3713 | -0.18 | 1.0 | 0.7390 | -0.06 | 1.0 | 0.3843 | -0.21 | 0.0 | 0.3460 | -0.17 | 0.0 | 0.0338 | -0.11 | 1.0 | 0.0325 | -0.18 | 0.0 | 0.0005 | -0.23 |
| | GL | | | | 0.0 | 0.1616 | -0.20 | 1.0 | 0.1337 | -0.06 | 1.0 | 0.0001 | -0.19 | 0.0 | 0.8187 | -0.12 | 0.0 | 0.6173 | -0.12 | 1.0 | 0.0093 | -0.19 | 0.0 | 0.0016 | -0.28 |
| HMM | US | 1.0 | 0.1337 | -0.09 | | | | 1.0 | 0.6833 | -0.19 | 1.0 | 0.0030 | -0.19 | 1.0 | 0.4671 | -0.12 | 1.0 | 0.6833 | -0.19 | 1.0 | 0.0515 | -0.22 | 1.0 | 0.0105 | -0.29 |
| | EU | 1.0 | 0.7056 | -0.19 | | | | 1.0 | 0.4655 | -0.19 | 1.0 | 0.0060 | -0.13 | 1.0 | 0.8349 | -0.16 | 1.0 | 0.8576 | -0.24 | 1.0 | 0.0191 | -0.19 | 1.0 | 0.0067 | -0.26 |
| | UK | 1.0 | 0.3713 | -0.13 | | | | 1.0 | 0.2515 | -0.19 | 1.0 | 0.0833 | -0.15 | 0.5 | 1.0000 | -0.11 | 1.0 | 0.6833 | -0.21 | 1.0 | 0.0030 | -0.13 | 1.0 | 0.0412 | -0.27 |
| | GL | 1.0 | 0.1616 | -0.12 | | | | 1.0 | 0.0279 | -0.15 | 1.0 | 0.0000 | -0.11 | 1.0 | 0.2009 | -0.11 | 1.0 | 0.3361 | -0.15 | 1.0 | 0.0002 | -0.14 | 1.0 | 0.0771 | -0.28 |
| FF (hmm init) | US | 1.0 | 0.2852 | -0.08 | 0.0 | 0.6833 | -0.22 | | | | 1.0 | 0.0131 | -0.25 | 1.0 | 0.8528 | -0.23 | 0.5 | 1.0000 | -0.10 | 1.0 | 0.1048 | -0.23 | 1.0 | 0.0009 | -0.27 |
| | EU | 0.0 | 0.5932 | -0.11 | 0.0 | 0.4655 | -0.24 | | | | 1.0 | 0.0704 | -0.22 | 0.0 | 0.6123 | -0.21 | 0.0 | 0.2254 | -0.15 | 1.0 | 0.1011 | -0.24 | 1.0 | 0.0016 | -0.32 |
| | UK | 0.0 | 0.7390 | -0.07 | 0.0 | 0.2515 | -0.18 | | | | 1.0 | 0.4930 | -0.22 | 0.0 | 0.2254 | -0.16 | 0.0 | 0.0348 | -0.13 | 1.0 | 0.0515 | -0.19 | 0.0 | 0.0010 | -0.27 |
| | GL | 0.0 | 0.1337 | -0.13 | 0.0 | 0.0279 | -0.28 | | | | 1.0 | 0.0013 | -0.18 | 0.0 | 0.1938 | -0.21 | 0.0 | 0.0881 | -0.17 | 1.0 | 0.0897 | -0.22 | 0.0 | 0.0001 | -0.33 |
| CNN (hmm init) | US | 0.0 | 0.0633 | -0.40 | 0.0 | 0.0030 | -0.39 | 0.0 | 0.0131 | -0.42 | | | | 0.0 | 0.0209 | -0.42 | 0.0 | 0.0131 | -0.42 | 0.0 | 0.2369 | -0.27 | 0.0 | 0.0000 | -0.51 |
| | EU | 0.0 | 0.0268 | -0.32 | 0.0 | 0.0060 | -0.30 | 0.0 | 0.0704 | -0.33 | | | | 0.0 | 0.0070 | -0.42 | 0.0 | 0.0095 | -0.36 | 0.0 | 0.8760 | -0.25 | 0.0 | 0.0000 | -0.43 |
| | UK | 0.0 | 0.3843 | -0.27 | 0.0 | 0.0833 | -0.25 | 0.0 | 0.4930 | -0.27 | | | | 0.0 | 0.1060 | -0.28 | 0.0 | 0.0705 | -0.34 | 0.0 | 0.1574 | -0.17 | 0.0 | 0.0018 | -0.39 |
| | GL | 0.0 | 0.0001 | -0.41 | 0.0 | 0.0000 | -0.41 | 0.0 | 0.0013 | -0.36 | | | | 0.0 | 0.0000 | -0.36 | 0.0 | 0.0000 | -0.39 | 0.0 | 0.0961 | -0.29 | 0.0 | 0.0000 | -0.49 |
| LSTM (hmm init) | US | 1.0 | 0.5318 | -0.16 | 0.0 | 0.4671 | -0.16 | 0.0 | 0.8528 | -0.25 | 1.0 | 0.0209 | -0.26 | | | | 0.0 | 0.8528 | -0.25 | 1.0 | 0.1282 | -0.21 | 0.0 | 0.0003 | -0.26 |
| | EU | 1.0 | 0.8619 | -0.23 | 0.0 | 0.8349 | -0.17 | 1.0 | 0.6123 | -0.23 | 1.0 | 0.0070 | -0.12 | | | | 0.0 | 0.7458 | -0.29 | 1.0 | 0.0196 | -0.16 | 0.0 | 0.0042 | -0.28 |
| | UK | 1.0 | 0.3460 | -0.11 | 0.5 | 1.0000 | -0.11 | 1.0 | 0.2254 | -0.10 | 1.0 | 0.1060 | -0.18 | | | | 0.0 | 0.6376 | -0.16 | 1.0 | 0.0030 | -0.13 | 0.0 | 0.0184 | -0.23 |
| | GL | 1.0 | 0.8187 | -0.11 | 0.0 | 0.2009 | -0.18 | 1.0 | 0.1938 | -0.14 | 1.0 | 0.0000 | -0.11 | | | | 0.0 | 0.8349 | -0.15 | 1.0 | 0.0023 | -0.12 | 0.0 | 0.0025 | -0.28 |
| FF (hmm init + reg) | US | 1.0 | 0.2060 | -0.05 | 0.0 | 0.6833 | -0.22 | 0.5 | 1.0000 | -0.10 | 1.0 | 0.0131 | -0.25 | 1.0 | 0.8528 | -0.23 | | | | 1.0 | 0.0956 | -0.22 | 0.0 | 0.0009 | -0.27 |
| | EU | 1.0 | 0.4056 | -0.07 | 1.0 | 0.8576 | -0.24 | 1.0 | 0.2254 | -0.09 | 1.0 | 0.0095 | -0.19 | 1.0 | 0.7458 | -0.27 | | | | 1.0 | 0.0236 | -0.25 | 0.0 | 0.0142 | -0.27 |
| | UK | 1.0 | 0.0338 | -0.02 | 1.0 | 0.6833 | -0.18 | 1.0 | 0.0348 | -0.03 | 1.0 | 0.0705 | -0.22 | 1.0 | 0.6376 | -0.16 | | | | 1.0 | 0.0030 | -0.18 | 0.0 | 0.0209 | -0.17 |
| | GL | 1.0 | 0.6173 | -0.09 | 1.0 | 0.3361 | -0.20 | 1.0 | 0.0881 | -0.14 | 1.0 | 0.0000 | -0.14 | 1.0 | 0.8349 | -0.14 | | | | 1.0 | 0.0025 | -0.15 | 0.0 | 0.0010 | -0.23 |
| CNN (hmm init + reg) | US | 0.0 | 0.3175 | -0.30 | 0.0 | 0.0515 | -0.36 | 0.0 | 0.1048 | -0.34 | 0.0 | 0.2369 | -0.20 | 0.0 | 0.1282 | -0.31 | 0.0 | 0.0956 | -0.33 | | | | 0.0 | 0.0000 | -0.40 |
| | EU | 0.0 | 0.0526 | -0.35 | 0.0 | 0.0191 | -0.34 | 0.0 | 0.1011 | -0.34 | 0.0 | 0.8760 | -0.25 | 0.0 | 0.0196 | -0.16 | 0.0 | 0.0236 | -0.40 | | | | 0.0 | 0.0000 | -0.43 |
| | UK | 0.0 | 0.0325 | -0.32 | 0.0 | 0.0030 | -0.32 | 0.0 | 0.0515 | -0.32 | 0.0 | 0.1574 | -0.25 | 0.0 | 0.0030 | -0.13 | 0.0 | 0.0030 | -0.38 | | | | 0.0 | 0.0000 | -0.43 |
| | GL | 0.0 | 0.0093 | -0.33 | 0.0 | 0.0002 | -0.35 | 0.0 | 0.0897 | -0.31 | 0.0 | 0.0961 | -0.20 | 0.0 | 0.0023 | -0.29 | 0.0 | 0.0025 | -0.32 | | | | 0.0 | 0.0000 | -0.40 |
| LSTM (hmm init + reg) | US | 1.0 | 0.0001 | -0.04 | 1.0 | 0.0105 | -0.11 | 1.0 | 0.0000 | -0.19 | 1.0 | 0.0000 | -0.17 | 1.0 | 0.0003 | -0.02 | 1.0 | 0.0000 | -0.09 | 1.0 | 0.0000 | -0.09 | | | |
| | EU | 1.0 | 0.0038 | -0.11 | 1.0 | 0.0067 | -0.09 | 1.0 | 0.0016 | -0.11 | 1.0 | 0.0000 | -0.13 | 1.0 | 0.0042 | -0.09 | 1.0 | 0.0142 | -0.11 | 1.0 | 0.0000 | -0.15 | | | |
| | UK | 1.0 | 0.0005 | -0.02 | 1.0 | 0.0412 | -0.13 | 1.0 | 0.0010 | -0.06 | 1.0 | 0.0018 | -0.17 | 1.0 | 0.0184 | -0.08 | 1.0 | 0.0209 | -0.04 | 1.0 | 0.0000 | -0.13 | | | |
| | GL | 1.0 | 0.0016 | -0.09 | 1.0 | 0.0771 | -0.17 | 1.0 | 0.0001 | -0.17 | 1.0 | 0.0000 | -0.20 | 1.0 | 0.0025 | -0.05 | 1.0 | 0.0010 | -0.08 | 1.0 | 0.0000 | -0.08 | | | |

Notes: This table displays the model comparison as a matrix. The models are listed per row (assessed model) and are compared to their peers which are listed per column (benchmark model). If a value indicate an advantage of one model over the other, it is colored green, otherwise red. P-values are colored if significant (< 0.1). Dominance (dom) values are coloured green is dominant (i.e. 0.00) and red if clearly not dominant (i.e. ≤ -0.1).

**Table 4, Panel D:**   Comparison of estimated 5% VaR of long bonds with and without balanced incentive function (i.e. regularized)

| | | Classic | | | HMM | | | FF (hmm init) | | | CNN (hmm init) | | | LSTM (hmm init) | | | FF (hmm init + reg) | | | CNN (hmm init + reg) | | | LSTM (hmm init + reg) | | |
|---|---|---|---|---|---|---|---|---|---|---|---|---|---|---|---|---|---|---|---|---|---|---|---|---|
| model | region | comp | pvalue | dom | comp | pvalue | dom | comp | pvalue | dom | comp | pvalue | dom | comp | pvalue | dom | comp | pvalue | dom | comp | pvalue | dom | comp | pvalue | dom |
| **Classic** | US | | | | 1.0 | 0.1968 | -0.09 | 1.0 | 0.0001 | 0.00 | 1.0 | 0.0000 | -0.09 | 1.0 | 0.0184 | -0.07 | 1.0 | 0.0046 | 0.00 | 1.0 | 0.0027 | -0.09 | 0.0 | 0.0028 | -0.29 |
| | EU | | | | 1.0 | 0.4388 | -0.09 | 1.0 | 0.4798 | -0.04 | 1.0 | 0.0004 | -0.04 | 1.0 | 0.0707 | -0.06 | 0.5 | 1.0000 | -0.04 | 1.0 | 0.0001 | -0.09 | 0.0 | 0.0075 | -0.17 |
| | UK | | | | 1.0 | 0.0833 | -0.06 | 1.0 | 0.0015 | 0.00 | 1.0 | 0.0000 | -0.04 | 1.0 | 0.1317 | -0.06 | 1.0 | 0.4797 | -0.06 | 1.0 | 0.0000 | -0.04 | 0.0 | 0.0009 | -0.22 |
| | GL | | | | 1.0 | 0.2254 | -0.11 | 1.0 | 0.0000 | -0.02 | 1.0 | 0.0124 | -0.05 | 1.0 | 0.0000 | -0.09 | 1.0 | 0.0008 | -0.02 | 1.0 | 0.0000 | -0.09 | 0.0 | 0.0348 | -0.16 |
| **HMM** | US | 0.0 | 0.1968 | -0.17 | | | | 1.0 | 0.0184 | -0.07 | 1.0 | 0.0002 | -0.07 | 1.0 | 0.1656 | -0.07 | 1.0 | 0.3175 | -0.05 | 1.0 | 0.0253 | -0.08 | 0.0 | 0.0001 | -0.33 |
| | EU | 0.0 | 0.4388 | -0.12 | | | | 0.0 | 0.7817 | -0.10 | 1.0 | 0.0017 | -0.04 | 1.0 | 0.3175 | -0.08 | 0.0 | 0.4671 | -0.14 | 0.0 | 0.0008 | -0.12 | 0.0 | 0.0045 | -0.24 |
| | UK | 0.0 | 0.0833 | -0.16 | | | | 1.0 | 0.2484 | -0.07 | 1.0 | 0.0000 | -0.02 | 1.0 | 0.7632 | -0.11 | 1.0 | 0.2060 | -0.12 | 1.0 | 0.0000 | -0.02 | 0.0 | 0.0001 | -0.32 |
| | GL | 0.0 | 0.2254 | -0.18 | | | | 1.0 | 0.0002 | -0.03 | 1.0 | 0.0000 | -0.02 | 1.0 | 0.2515 | -0.12 | 1.0 | 0.0593 | -0.08 | 1.0 | 0.0006 | -0.10 | 0.0 | 0.0105 | -0.28 |
| **FF (hmm init)** | US | 0.0 | 0.0001 | -0.21 | 0.0 | 0.0184 | -0.20 | | | | 1.0 | 0.0833 | -0.13 | 0.0 | 0.2754 | -0.19 | 0.0 | 0.0522 | -0.14 | 0.5 | 1.0000 | -0.20 | 0.0 | 0.0000 | -0.43 |
| | EU | 0.0 | 0.4798 | -0.07 | 0.0 | 0.7817 | -0.08 | | | | 1.0 | 0.0017 | -0.06 | 1.0 | 0.2254 | -0.08 | 0.0 | 0.4145 | -0.06 | 0.0 | 0.0003 | -0.11 | 0.0 | 0.0027 | -0.20 |
| | UK | 0.0 | 0.0015 | -0.17 | 0.0 | 0.2484 | -0.13 | | | | 1.0 | 0.0000 | -0.05 | 1.0 | 0.1656 | -0.15 | 0.0 | 0.0209 | -0.17 | 0.0 | 0.0053 | -0.12 | 0.0 | 0.0000 | -0.35 |
| | GL | 0.0 | 0.0000 | -0.29 | 0.0 | 0.0002 | -0.25 | | | | 1.0 | 0.0028 | -0.05 | 1.0 | 0.0185 | -0.25 | 0.0 | 0.0066 | -0.13 | 1.0 | 0.7317 | -0.21 | 0.0 | 0.0000 | -0.40 |
| **CNN (hmm init)** | US | 0.0 | 0.0000 | -0.37 | 0.0 | 0.0002 | -0.29 | 0.0 | 0.0833 | -0.23 | | | | 0.0 | 0.0105 | -0.28 | 0.0 | 0.0025 | -0.28 | 0.0 | 0.0605 | -0.20 | 0.0 | 0.0000 | -0.48 |
| | EU | 0.0 | 0.0004 | -0.23 | 0.0 | 0.0017 | -0.20 | 0.0 | 0.0017 | -0.22 | | | | 0.0 | 0.0412 | -0.20 | 0.0 | 0.0007 | -0.24 | 0.0 | 0.1938 | -0.13 | 0.0 | 0.0000 | -0.35 |
| | UK | 0.0 | 0.0000 | -0.41 | 0.0 | 0.0000 | -0.33 | 0.0 | 0.0000 | -0.30 | | | | 0.0 | 0.0000 | -0.38 | 0.0 | 0.0000 | -0.38 | 0.0 | 0.1267 | -0.17 | 0.0 | 0.0000 | -0.54 |
| | GL | 0.0 | 0.0000 | -0.41 | 0.0 | 0.0000 | -0.35 | 0.0 | 0.0028 | -0.20 | | | | 0.0 | 0.0000 | -0.27 | 0.0 | 0.0000 | -0.34 | 0.0 | 0.0185 | -0.21 | 0.0 | 0.0000 | -0.51 |
| **LSTM (hmm init)** | US | 0.0 | 0.0184 | -0.22 | 1.0 | 0.1656 | -0.14 | 1.0 | 0.2754 | -0.12 | 1.0 | 0.0105 | -0.12 | | | | 0.0 | 0.6173 | -0.14 | 1.0 | 0.3361 | -0.17 | 0.0 | 0.0000 | -0.35 |
| | EU | 0.0 | 0.0707 | -0.14 | 1.0 | 0.3175 | -0.13 | 0.0 | 0.2254 | -0.14 | 1.0 | 0.0412 | -0.09 | | | | 0.0 | 0.0896 | -0.16 | 1.0 | 0.0030 | -0.11 | 0.0 | 0.0001 | -0.24 |
| | UK | 0.0 | 0.1317 | -0.15 | 1.0 | 0.7632 | -0.09 | 1.0 | 0.1656 | -0.07 | 1.0 | 0.0000 | -0.07 | | | | 1.0 | 0.4388 | -0.16 | 1.0 | 0.0002 | -0.07 | 0.0 | 0.0001 | -0.29 |
| | GL | 0.0 | 0.0124 | -0.20 | 1.0 | 0.2515 | -0.18 | 1.0 | 0.0185 | -0.11 | 1.0 | 0.0000 | -0.08 | | | | 1.0 | 0.5129 | -0.14 | 1.0 | 0.0105 | -0.12 | 0.0 | 0.0001 | -0.28 |
| **FF (hmm init + reg)** | US | 0.0 | 0.0046 | -0.13 | 0.0 | 0.3175 | -0.10 | 1.0 | 0.0522 | -0.05 | 1.0 | 0.0025 | -0.06 | 1.0 | 0.6173 | -0.07 | | | | 1.0 | 0.1616 | -0.14 | 0.0 | 0.0000 | -0.35 |
| | EU | 0.5 | 1.0000 | -0.04 | 0.0 | 0.4671 | -0.10 | 1.0 | 0.4145 | -0.03 | 1.0 | 0.0007 | -0.06 | 1.0 | 0.0896 | -0.07 | | | | 1.0 | 0.0001 | -0.09 | 0.0 | 0.0124 | -0.19 |
| | UK | 1.0 | 0.4797 | -0.10 | 1.0 | 0.2060 | -0.06 | 1.0 | 0.0209 | -0.04 | 1.0 | 0.0000 | -0.04 | 1.0 | 0.4388 | -0.12 | | | | 1.0 | 0.0000 | -0.04 | 0.0 | 0.0008 | -0.27 |
| | GL | 0.0 | 0.0008 | -0.21 | 1.0 | 0.0593 | -0.08 | 1.0 | 0.0066 | -0.05 | 1.0 | 0.0000 | -0.03 | 1.0 | 0.5129 | -0.18 | | | | 1.0 | 0.0482 | -0.15 | 0.0 | 0.0000 | -0.32 |
| **CNN (hmm init + reg)** | US | 0.0 | 0.0027 | -0.29 | 0.0 | 0.0253 | -0.21 | 0.5 | 1.0000 | -0.20 | 0.0 | 0.0605 | -0.10 | 0.0 | 0.3361 | -0.23 | 0.0 | 0.1616 | -0.23 | | | | 0.0 | 0.0000 | -0.43 |
| | EU | 0.0 | 0.0001 | -0.32 | 0.0 | 0.0008 | -0.32 | 0.0 | 0.0003 | -0.32 | 0.0 | 0.1938 | -0.19 | 0.0 | 0.0030 | -0.27 | 0.0 | 0.0001 | -0.32 | | | | 0.0 | 0.0000 | -0.40 |
| | UK | 0.0 | 0.0000 | -0.36 | 0.0 | 0.0000 | -0.27 | 0.0 | 0.0053 | -0.29 | 0.0 | 0.1267 | -0.17 | 0.0 | 0.0002 | -0.32 | 0.0 | 0.0000 | -0.33 | | | | 0.0 | 0.0000 | -0.49 |
| | GL | 0.0 | 0.0000 | -0.37 | 0.0 | 0.0006 | -0.32 | 1.0 | 0.7317 | -0.23 | 0.0 | 0.0185 | -0.09 | 0.0 | 0.0105 | -0.28 | 0.0 | 0.0482 | -0.27 | | | | 0.0 | 0.0000 | -0.42 |
| **LSTM (hmm init + reg)** | US | 1.0 | 0.0028 | -0.07 | 1.0 | 0.0001 | -0.05 | 1.0 | 0.0000 | -0.05 | 1.0 | 0.0000 | -0.02 | 1.0 | 0.0000 | 0.00 | 1.0 | 0.0000 | -0.02 | 1.0 | 0.0000 | -0.05 | | | |
| | EU | 1.0 | 0.0075 | -0.03 | 1.0 | 0.0045 | -0.07 | 1.0 | 0.0027 | -0.03 | 1.0 | 0.0000 | -0.05 | 1.0 | 0.0001 | -0.02 | 1.0 | 0.0124 | -0.05 | 1.0 | 0.0000 | -0.05 | | | |
| | UK | 1.0 | 0.0009 | 0.00 | 1.0 | 0.0001 | -0.03 | 1.0 | 0.0000 | -0.03 | 1.0 | 0.0000 | -0.03 | 1.0 | 0.0001 | -0.04 | 1.0 | 0.0008 | -0.03 | 1.0 | 0.0000 | -0.03 | | | |
| | GL | 1.0 | 0.0348 | -0.04 | 1.0 | 0.0105 | -0.10 | 1.0 | 0.0000 | -0.04 | 1.0 | 0.0000 | -0.06 | 1.0 | 0.0001 | -0.02 | 1.0 | 0.0001 | -0.04 | 1.0 | 0.0000 | -0.04 | | | |

Notes: This table displays the model comparison as a matrix. The models are listed per row (assessed model) and are compared to their peers which are listed per column (benchmark model). If a value indicate an advantage of one model over the other, it is colored green, otherwise red. P-values are colored if significant (< 0.1). Dominance (dom) values are coloured green is dominant (i.e. 0.00) and red if clearly not dominant (i.e. ≤ -0.1).

Table 5: Monetary costs of 5% VaR breaches in equities (above) and long bonds (below)

| region | US | | EU | | UK | | GL | |
|---|---|---|---|---|---|---|---|---|
| | acc. loss per year | avg loss per breach | acc. loss per year | avg loss per breach | acc. loss per year | avg loss per breach | acc. loss per year | avg loss per breach |
| **model** | | | | | | | | |
| Classic | -5.27% | -0.68% | -7.33% | -0.67% | -5.63% | -0.74% | -6.41% | -0.76% |
| HMM | -4.74% | -0.50% | -6.93% | -0.49% | -5.18% | -0.54% | -5.50% | -0.51% |
| FF (hmm init) | -5.29% | -0.68% | -7.27% | -0.69% | -5.85% | -0.72% | -6.43% | -0.76% |
| CNN (hmm init) | -5.54% | -0.80% | -7.32% | -0.81% | -5.50% | -0.78% | -6.91% | -0.97% |
| LSTM (hmm init) | -5.11% | -0.61% | -7.25% | -0.56% | -5.39% | -0.60% | -6.15% | -0.71% |
| FF (hmm init + reg) | -5.02% | -0.56% | -6.85% | -0.54% | -5.59% | -0.66% | -6.10% | -0.66% |
| CNN (hmm init + reg) | -5.38% | -0.75% | -7.91% | -0.91% | -6.23% | -1.07% | -7.58% | -1.02% |
| LSTM (hmm init + reg) | -4.30% | -0.18% | -6.05% | -0.10% | -4.72% | -0.32% | -5.10% | -0.35% |
| region | US | | EU | | UK | | GL | |
| | acc. loss per year | avg loss per breach | acc. loss per year | avg loss per breach | acc. loss per year | avg loss per breach | acc. loss per year | avg loss per breach |
| **model** | | | | | | | | |
| Classic | -1.67% | -0.33% | -1.53% | -0.24% | -1.21% | -0.20% | -1.61% | -0.28% |
| HMM | -1.66% | -0.32% | -1.52% | -0.24% | -1.23% | -0.22% | -1.59% | -0.25% |
| FF (hmm init) | -1.87% | -0.41% | -1.61% | -0.26% | -1.37% | -0.27% | -1.99% | -0.41% |
| CNN (hmm init) | -2.05% | -0.47% | -1.86% | -0.33% | -1.55% | -0.36% | -2.13% | -0.47% |
| LSTM (hmm init) | -1.81% | -0.37% | -1.66% | -0.27% | -1.30% | -0.22% | -1.71% | -0.32% |
| FF (hmm init + reg) | -1.75% | -0.36% | -1.52% | -0.23% | -1.24% | -0.22% | -1.81% | -0.35% |
| CNN (hmm init + reg) | -1.88% | -0.40% | -1.85% | -0.33% | -1.53% | -0.33% | -1.99% | -0.41% |
| LSTM (hmm init + reg) | -1.27% | -0.12% | -1.29% | -0.16% | -0.98% | -0.02% | -1.33% | -0.13% |

Notes: This table displays all losses that exceed the estimated VaR threshold of a model. Two ways of aggregation are given: accumulated loss per year and average loss per breach as calculated from given backtests per region.

Table 6: Comparison (comp) values of all three types of neural networks in all four regions for 2,000 vs. 1,000 training days

| Training Days | 2,000 days | 1,000 days |
|---|---|---|
| Var 1% Equities | 11 | 1 |
| Var 1% Long Bonds | 10.5 | 1.5 |
| Var 5% Equities | 12 | 0 |
| Var 5% Long Bonds | 12 | 0 |

Notes: This table compared all three types of neural networks in all four regions (12 comparisons in total) based on the network being fed with 2,000 or 1,000 training days. The comp values are computed equivalent to those in Table 4. The total of each row has to be 12 given the 12 underlying comparisons.